\documentclass[11pt]{article}

\usepackage{fullpage}

\usepackage{amsfonts,color,amssymb,graphicx,stmaryrd}
\usepackage{amsmath}
\usepackage{url}

\newtheorem{theorem}{Theorem}

\newtheorem{proposition}[theorem]{Proposition}

\def\Proof{{\bf Proof.}}
\def\lt{\langle}
\def\rt{\rangle}
\def\lsem{\lt\!\lt}
\def\rsem{\rt\!\rt}

\def\W{W}

\def\H{H}
\def\BT{B}
\def\Proof{{\bf Proof.}}

\def\beginstring{\lhd}
\newcommand{\TNW}[1]{\lsem{#1}\rsem}
\def\E{E}
\def\S{S}
\def\pms{\Pi}
\def\pm{\pi}
\def\A{{\cal A}}

\def\val{\alpha}
\def\pval{\beta}
\def\sval{\Lambda}
\def\occ{\varphi}
\def\hole{?}

\newcommand{\brac}[1]{\ensuremath{\llbracket #1\rrbracket}}
\def\mypar#1{\vspace*{.1in}\noindent{\bf {#1}:}}
\def\bigO{{O}}
\def\qed{{\bf $\Box$}}
\def\rev{\mbox{{\it reverse}}}
\def\nwbt{\mbox{{\it nw\_bt}}}
\def\btnw{\mbox{{\it bt\_nw}}}
\def\sep{\,|\,}
\def\matchrel{\rightsquigarrow}
\def\comment#1{}
\def\zero{{\bf 0}}
\def\dom{\mbox{\it Dom\/}}
\def\img{\mbox{\it Img\/}}
\def\preimg{\mbox{\it PreImg\/}}
\def\wc{\underline{\hspace{0.2cm}}}

\def\lwm{\mbox{{\sc lwm}}}
\def\lc{\mbox{{\sc lc}}}
\def\lwms{\mbox{{\sc wms}}}
\def\ret{\mbox{{\sc ret}}}

\begin{document}

\title{Streaming Tree Transducers}

\author{Rajeev Alur and Loris D'Antoni\\
University of Pennsylvania}

\maketitle

\begin{abstract}
Theory of tree transducers provides a foundation for understanding expressiveness
and complexity of analysis problems for specification languages for transforming
hierarchically structured data such as XML documents.
We introduce {\em streaming tree transducers} as an analyzable,
executable,  and expressive model
for transforming unranked ordered trees (and hedges) in a single pass.
Given a linear encoding of the input tree, the transducer makes a single left-to-right
pass through the input, and computes the output in linear time
using a finite-state control, a visibly pushdown stack, and
a finite number of variables that store output chunks that can be combined using
the operations of string-concatenation and tree-insertion.
We prove that the expressiveness of the model coincides with transductions definable using
monadic second-order logic (MSO).
Existing models of tree transducers either cannot implement all MSO-definable transformations,
or require {\em regular look ahead} that prohibits single-pass implementation.
We show a variety of analysis problems such as  {\em type-checking\/} and
checking {\em functional equivalence\/}  are decidable for our model.
\end{abstract}

\section{Introduction}
Finite-state machines and logics for specifying tree transformations
offer a suitable theoretical foundation for studying
expressiveness and complexity of analysis problems for
languages for processing and transforming XML documents.
Representative formalisms for specifying tree transductions include
finite-state top-down and bottom-up tree transducers,
Macro tree transducers (MTT),
attribute grammars,
MSO (monadic second-order logic) definable graph transductions,
and specialized programming languages such as {\sc XSLT} and
{\sc XDuce}~\cite{TATA,Cou94,EM99,EV85,MSV02,HP03,MN05b,Ho11}.

In this paper, we propose the model of {\em streaming tree transducers\/} (STT)
which has the following three properties:
(1)~{\em Single-pass linear-time processing:} an STT is a deterministic machine
that computes the output using a single left-to-right pass
through the linear encoding of the input tree processing each symbol in constant time;
(2)~{\em Expressiveness:} STTs specify exactly the class of MSO-definable transductions; and
(3)~{\em Analyzability:} decision problems such as type checking and
checking functional equivalence of two STTs, are decidable.
The last two features indicate that our model has the commonly accepted trade-off
between analyzability and expressiveness in formal language theory.
The motivation for designing streaming algorithms that can process a document in a single
pass has led to streaming models for
checking membership in a regular tree language and for querying \cite{SV02,NS02,MSV02,MV09},
but there is no previous model that can compute all MSO-definable transformations in a single pass
(see Section~6 for detailed comparisons of STTs with prior models).

The transducer model integrates features of {\em visibly pushdown automata\/}, equivalently  {\em nested word
automata}~\cite{AM09}, and {\em streaming string transducers\/}~\cite{AC10,AC11}.
In our model, the input tree is encoded as a {\em nested word\/}, which is a string
over alphabet symbols, tagged with open/close brackets (or equivalently, call/return types)
to indicate the hierarchical structure~\cite{SV02,AM09}.
The streaming tree transducer reads the input nested word
left-to-right in a single pass.
It uses finitely many states,
together with a stack, but the type of operation applied to the  stack
at each step is determined by the hierarchical structure of the tags in the input.
The output is computed using
a finite set of variables that range over output nested words, possibly with {\em holes\/}
that are used as place-holders for inserting subtrees.
At each step, the transducer reads the next symbol of the input.
If the symbol is an internal symbol, then the transducer updates its state and
the output variables.
If the symbol is a call symbol, then the transducer pushes a stack symbol, along with
updated values of variables, updates the state, and reinitializes the variables.
While processing a return symbol, the stack is popped, and the new state and new
values for the variables are determined using the current state, current variables,
popped symbol, and popped values from the stack.
In each type of transition,
the variables are updated using expressions that allow adding new symbols,
{\em string concatenation\/}, and {\em tree insertion\/} (simulated by replacing the hole
with another expression).
A key restriction is that variables are updated in a manner that ensures that each value can
contribute at most once to the eventual output, without duplication.
This {\em single-use-restriction\/} is enforced via a binary {\em conflict\/} relation over variables:
no output term combines conflicting variables, and variable occurrences in right-hand sides
during each update are consistent with the conflict relation.
The transformation computed by the model can be implemented as a single-pass linear-time algorithm.

To understand the novel features of our model, let us consider two kinds of transformations.
First, suppose we want to select and output the sequence of all subtrees that match
a pattern, that is specified by a regular query over the entire input,
and not just the prefix read so far.
To implement this query, the transducer uses multiple variables to store alternative outputs,
and exploiting regularity to maintain only a bounded number of choices at each step.
In contrast, for existing transducer models, either {\em regular look ahead\/}
(that is, allowing the transducer to make
decisions based on a regular property of the suffix of the input it has not yet seen)
is essential
to define such a transduction, thereby necessitating a preprocessing pass over the input
(for example, MTTs with single use restriction),
or in absence of regular look ahead, a direct implementation of the operational semantics leads
to exponential growth in the size of intermediate derivations with the length of the input
(for example, MTTs and MTTs with weak finite copying restriction).
Second, suppose the transformation requires {\em swapping\/} of subtrees.
The operations of concatenation and tree-insertion allows an STT to
implement this transformation easily.
This ability to combine  previously computed answers seems to be missing from existing transducer models.
We illustrate the proposed model using examples such as reverse, swap, tag-based sorting,
that the natural single-pass linear-time algorithms for implementing these transformations
correspond to STTs.

We show that the model can be simplified in natural ways if we want to restrict either
the input or the output, to either strings or ranked trees.
For example, to compute transformations that output strings it suffices to consider
variable updates that allow only concatenation, and to compute transformations that output ranked trees
it suffices to consider variable updates that allow only tree insertion.
The restriction to the case of ranked trees as inputs gives the model of {\em bottom-up ranked-tree
transducers\/}. As far as we know, this is the only transducer model that processes trees in a
bottom-up manner, and can compute all MSO-definable transformations.

The main technical result in the paper is that the class of transductions definable using streaming tree transducers
is exactly the class of MSO-definable transductions.
The starting point for our result is  the known equivalence of MSO-definable transductions and
Macro Tree Transducers with regular look-ahead and single-use restriction,
over ranked trees~\cite{EM99}.
Our proof proceeds by establishing two key properties of STTs:
the model is closed under {\em regular look ahead} and under
{\em functional composition}.
These proofs are challenging due to  the requirement that
a transducer can use only a fixed number of variables
that can be updated by assignments that obey the single-use-restriction rules,
and we develop them in a modular fashion by introducing
intermediate results (for example, we establish that allowing variables to range over
trees that contain multiple parameters that can be selectively substituted during updates,
does not increase expressiveness).

We show a variety of analysis questions for our transducer model to be decidable.
Given a regular language $L_1$ of input trees and a regular language $L_2$ of output trees,
the {\em type checking\/} problem is to determine if the output of
the transducer on an input in $L_1$ is guaranteed to be in $L_2$.
We establish an {\sc Exptime\/} upper bound on type checking.
For checking functional equivalence of two streaming tree transducers,
we show that if the two transducers are inequivalent, then we can construct a
pushdown automaton $A$ over the alphabet $\{0,1\}$ such that $A$ accepts a word with
equal number of $0$'s and $1$'s exactly when there is an input on which the two transducers compute different
outputs.
Using known techniques for computing the Parikh images of context-free languages~\cite{EM06,SSMH04,Esp97},
this leads to a {\sc NExpTime} upper bound for checking functional inequivalence of two STTs.
Assuming a bounded number of  variables, the upper bound on the parametric complexity becomes {\sc NP}.
Improving the {\sc NExpTime} bound remains a challenging open problem.

\section{Transducer Model}
\subsection{Preliminaries}

\mypar{Nested  Words}
Data with both linear and hierarchical structure can be encoded using
nested words~\cite{AM09}.
Given a set $\Sigma$ of symbols, the {\em tagged alphabet\/}
$\hat\Sigma$  consists of the symbols $a$, $\lt a$, and $a \rt$, for
each $a\in \Sigma$.
A {\em nested word\/} over $\Sigma$ is a finite sequence over $\hat\Sigma$.
For a nested word $a_1\cdots a_k$, a position $j$, for $1\le j\le k$,
 is said to be a {\em call\/} position if the symbol $a_j$ is of the form $\lt a$,
a {\em return\/} position if the symbol $a_j$ is of the form $a \rt$,
and an {\em internal\/} position otherwise.
The tags induce a natural matching relation between call and return positions,
and in this paper, we are interested only in {\em well-matched\/} nested words
in which all calls/returns have matching returns/calls.
A string over $\Sigma$ is a nested word with only internal positions.
Nested words naturally encode ordered trees.
The empty tree is encoded by the empty string $\varepsilon$.
The tree with $a$-labeled root with subtrees $t_1,\ldots t_k$ as children,
in that order, is encoded by the nested word
$\lt a\, \TNW{t_1}\, \cdots  \TNW{t_k}\,a\rt$, where $\TNW{t_i}$ is the encoding of the subtree $t_i$.
This transformation can be viewed as an inorder traversal of the tree.
The encoding extends to {\em hedges} also: the encoding of a hedge is obtained
by concatenating the encodings of the trees it contains.
An $a$-labeled leaf corresponds to the nested word $\lt a a\rt$, we will
use $\lt a \rt$ as its abbreviation.
Thus, a binary tree with $a$-labeled root whose left-child is an $a$-labeled leaf and right-child is a $b$-labeled
leaf is encoded by the string $\lt a\,\lt a \rt\,\lt b \rt\,\ a\rt$.

\mypar{Nested Words with Holes}
A key operation that our transducer model relies on is {\em insertion\/}
of one nested word within another. In order to define this, we
consider nested words with holes, where a hole is represented by the special
symbol $\hole$.
For example,
the nested word $\lt a\,\hole\,\lt b \rt\,\ a\rt$ represents
an incomplete tree with $a$-labeled root whose right-child
is a $b$-labeled leaf such that the tree can be completed by adding a nested word to the
left of this leaf.
We require that a nested word can contain at most one hole, and we
use a binary type to keep track of whether a nested word contains a hole or not.
A type-0 nested word does not contain any holes, while a type-1 nested word contains one hole.
We can view a type-1 nested word as a unary function from nested words to nested words.
The set $\W_0(\Sigma)$ of type-0 nested words over the alphabet $\Sigma$ is defined by the grammar
\[W_0\ :=\ \varepsilon \sep a\sep\lt a\, W_0\, b\rt\sep W_0\, W_0,\]
for $a,b\in\Sigma$.
The set $\W_1(\Sigma)$ of type-1 nested words over the alphabet $\Sigma$
is defined by the grammar
\[W_1\ :=\ \hole \sep \lt a\, W_1\, b\rt \sep W_1\, W_0 \sep W_0\, W_1,\]
for $a,b\in\Sigma$.
A {\em nested-word language\/} over $\Sigma$ is a subset $L$ of $\W_0(\Sigma)$,
and a {\em nested-word transduction\/} from an input alphabet $\Sigma$ to an output alphabet $\Gamma$
is a {\em partial\/} function $f$ from $\W_0(\Sigma)$ to $\W_0(\Gamma)$.

\mypar{Nested  Word Expressions}
In our transducer model, the machine maintains a set of variables that range
over output nested words with holes.
Each variable has an associated binary type: a type-$k$ variable has type-$k$ nested words as values,
for $k=0,1$.
The variables are updated using typed expressions, where variables can appear on the right-hand side,
and we also allow substitution of the hole symbol by another expression.
Formally, a set $X$ of typed variables is a set that is partitioned into two sets $X_0$ and $X_1$
corresponding to the type-0 and type-1 variables.
Given an alphabet $\Sigma$ and a set $X$ of typed variables,
a {\em valuation\/} $\val$ is a function that maps $X_0$ to $\W_0(\Sigma)$
and $X_1$ to $\W_1(\Sigma)$.
Given an alphabet $\Sigma$ and a set $X$ of typed variables,
we define the sets $\E_k(X,\Sigma)$, for $k=0,1$, of type-$k$ expressions
by the grammars:
\begin{eqnarray*}
E_0 & := & \varepsilon \sep a \sep x_0 \sep \lt a\, E_0\, b\rt \sep E_0\, E_0  \sep E_1[E_0]\\
E_1 & := & \hole\sep x_1 \sep \lt a\, E_1\, b\rt \sep E_0\, E_1 \sep E_1\,E_0  \sep E_1[E_1],
\end{eqnarray*}
where $a,b\in\Sigma$, $x_0\in X_0$ and $x_1\in X_1$.
The clause $e[e']$ corresponds to substitution of the hole in a type-$1$ expression $e$
by another expression $e'$.
A valuation $\val$ for the variables $X$ naturally extends to a type-consistent function
that maps the expressions $\E_k(X,\Sigma)$ to values in $\W_k(\Sigma)$, for $k=0,1$.
Given an expression $e$, $\val(e)$ is obtained by replacing each variable $x$ by $\val(x)$,
and applying the substitution:
in particular, $\val(e[e'])$ is obtained by replacing the symbol $\hole$
in the type-1 nested word $\val(e)$ by the nested word $\val(e')$.

\mypar{Single Use Restriction}
The transducer updates variables $X$ using type-consistent assignments.
To achieve the desired expressiveness, we need to restrict the reuse of variables in right-hand sides.
In particular, we want to disallow the assignment $x := xx$ (which would double the length of $x$),
but allow the assignment $(x,y):=(x,x)$, provided the variables $x$ and $y$ are guaranteed
not to be combined later. For this purpose,
we assume that the set $X$ of variables is equipped with a binary relation
$\eta$: if $\eta(x, y)$, then $x$ and $y$ cannot be combined.
This ``conflict'' relation is required to be reflexive and symmetric (but need not be transitive).
Two conflicting variables cannot occur in the same expression used in the right-hand side of an update
or as output.
During an update, two conflicting variables
can occur in multiple right-hand sides for updating conflicting variables.
Thus, the assignment $(x,y):=(\lt a\, x a\rt[y],a\hole)$
is allowed, provided $\eta(x,y)$ does not hold;
the assignment $(x,y):=(ax[y],y)$ is not allowed; and
the assignment $(x,y):=(ax, x[b])$ is allowed, provided $\eta(x,y)$ holds.
Formally, given a set $X$ of typed variables with a reflexive symmetric binary conflict relation $\eta$, and an alphabet $\Sigma$,
an expression $e$ in $E(X,\Sigma)$ is said to be {\em consistent\/} with $\eta$, if
(1) each variable $x$ occurs at most once in $e$, and (2)
if $\eta(x,y)$ holds, then $e$ does not contain both $x$ and $y$.
Given sets $X$ and $Y$ of typed variables, a conflict relation $\eta$, and an alphabet $\Sigma$,
a {\em single-use-restricted assignment\/} is a function $\rho$ that maps each
type-$k$ variable $x$ in $X$ to a right-hand side expression in $\E_k(Y,\Sigma)$, for $k=0,1$, such that
(1) each expression $\rho(x)$ is consistent with $\eta$, and
(2) if $\eta(x,y)$ holds, and $\rho(x')$ contains $x$, and $\rho(y')$ contains $y$, then $\eta(x',y')$ must hold.
The set of such single-use-restricted assignments is denoted $\A(X,Y,\eta,\Sigma)$.

At a return, the transducer assigns the values to its variables $X$ using the
values popped from the stack as well as the values returned. For each variable $x$, we will use $x_p$ to
refer to the popped value of $x$.
Thus, each variable $x$ is updated using an expression over the variables $X\cup X_p$.
The conflict relation $\eta$ extends naturally to variables in $X_p$:
$\eta(x_p,y_p)$ holds exactly when $\eta(x,y)$ holds.
Then, the update at a return is specified by assignments in $\A(X,X\cup X_p,\eta,\Sigma)$.

When the conflict relation $\eta$ is the purely reflexive relation $\{(x,x)\mid x\in X\}$, the single-use-restriction means
that a variable $x$ can appear at most once in at most one right-hand side.
We refer to this special case as ``copyless''.

\subsection{Transducer Definition}
A streaming tree transducer is a deterministic machine that
reads the input nested word
left-to-right in a single pass.
It uses finitely many states,
together with a stack.
The use of the stack is dictated by
the hierarchical structure of the call/return tags in the input.
The output is computed using
a finite set of typed variables, with a conflict relation that restricts which variables can be combined,
that range over nested words
and the stack can be used to store values of these variables.
At each step, the transducer reads the next symbol of the input.
If the symbol is an internal symbol, then the transducer updates its state and
the nested-word variables.
If the symbol is a call symbol, then the transducer pushes a stack symbol,
updates the state,  stores updated values of variables in the stack, and reinitializes the variables.
While processing a return symbol, the stack is popped, and the new state and new
values for the variables are determined using the current state, current variables,
popped symbol, and popped variables from the stack.
In each type of transition,
the variables are updated in parallel using assignments in which
the right-hand sides are nested-word expressions.
We require that the update is type-consistent, and meets the single-use-restriction  with respect to
the conflict relation.
When the transducer consumes the entire input word,
the output nested word is produced by an expression that is consistent with the conflict relation.
These requirements ensure that at every step, at most one
copy of any value is contributed to the final output.

\mypar{STT syntax}
A \emph{deterministic streaming tree transducer} (STT) $\S$
from input alphabet $\Sigma$ to output alphabet $\Gamma$ consists of
a finite set of states $Q$;
a finite set of stack symbols $P$;
an initial state $q_0\in Q$;
a finite set of typed variables $X$ with a reflexive symmetric binary conflict relation $\eta$;
a partial output function $F : Q \mapsto \E_0(X,\Gamma)$ such that
each expression $F(q)$ is consistent with $\eta$;
an internal state-transition function $\delta_i:Q\times \Sigma\mapsto Q$;
a call state-transition function $\delta_c:Q\times \Sigma\mapsto Q\times P$;
a return state-transition function $\delta_r:Q\times P\times \Sigma\mapsto Q$;
an  internal variable-update function $\rho_i:Q\times \Sigma\mapsto \A(X,X,\eta,\Gamma)$;
a call variable-update function $\rho_c:Q\times \Sigma\mapsto \A(X,X,\eta,\Gamma)$;
and a return variable-update function $\rho_r:Q\times P\times \Sigma\mapsto \A(X,X\cup X_p,\eta,\Gamma)$.

\mypar{STT semantics}
To define the semantics of a streaming tree transducer, we consider configurations of the
form $(q,\sval,\val)$, where $q\in Q$ is a state,
$\val$ is a type-consistent valuation from variables $X$ to typed nested words over $\Gamma$, and
$\sval$ is a sequence of pairs $(p,\pval)$
such that $p\in P$ is a stack symbol and $\pval$ is a type-consistent valuation
from variables in $X$ to typed nested words over $\Gamma$.
The initial configuration is $(q_0,\varepsilon,\val_0)$
where $\val_0$ maps each type-0 variable to $\varepsilon$ and each type-1 variable to $\hole$.
The transition function $\delta$ over configurations is defined by:
\begin{enumerate}
\item {\bf Internal transitions:}
          $\delta((q,\sval,\val),a)=(\delta_i(q,a),\sval,\val\cdot\rho_i(q,a))$.
\item {\bf Call transitions:}
          $\delta((q,\sval,\val),\lt a)=(q',(p,\val\cdot\rho_c(q,a))\sval,\val_0)$, where
          $\delta_c(q,a)=(q',p)$.
\item {\bf Return transitions:}
          $\delta((q,(p,\pval)\sval,\val), a\rt)=(\delta_r(q,p,a),\sval,\val\cdot\pval_p\cdot\rho_r(q,p,a))$,
 where $\beta_p$ is the valuation for variables $X_p$ defined by $\beta_p(x_p)=\beta(x)$ for $x\in X$.
\end{enumerate}

For an input word $w\in\W_0(\Sigma)$,
if $\delta^*((q_0,\varepsilon,\val_0),w)=(q,\varepsilon,\val)$
then if $F(q)$ is undefined then so is $\brac{S}(w)$,
otherwise $\brac{S}(w)=\val(F(q))$.
We say that a nested word transduction $f$ from input alphabet $\Sigma$ to output alphabet $\Gamma$
is {\em STT-definable\/} if there exists an STT $\S$ such that $\brac{S}=f$.

An STT $S$ with variables $X$ is called {\em copyless\/} if the conflict relation $\eta$ equals
$\{(x,x)\mid x \in X\}$.

\subsection{Examples}

Streaming tree transducers can easily implement standard tree-edit operations such as
insertion, deletion, and relabeling.
We illustrate the interesting features of our model
using operations such as reverse, swap, and sorting based on fixed number of tags.
In each of these cases, the transducer mirrors the natural algorithm for implementing
the desired operation in a single pass.
In each example, the STT is copyless.

\mypar{Reverse}
Given a nested word $a_1a_2\cdots a_k$, its {\em reverse} is the nested word
$b_k \cdots b_2 b_1$, where for each $1\le j\le k$,
$b_j=a_j$ if $a_j$ is an internal symbol,
$b_j=\lt a$ if $a_j$ is a return symbol $a\rt$, and
$b_j=a \rt$ if $a_j$ is a call symbol $\lt a$.
As a tree transformation, {\em reverse} corresponds to recursively reversing the order of children at each node:
the reverse of $\lt a\,\lt b\, \lt d\rt\,\lt e\rt \,b\rt \,\lt c \rt \,a\rt$ is
$\lt a\,\lt c \rt\,\lt b\,\lt e\rt\, \lt d\rt\,b\rt\,a\rt$.
This transduction can be implemented
by a streaming tree transducer with a single state, a single type-0 variable $x$, and
stack symbols $\Sigma$:
the internal transition on input $a$ updates $x$ to $a\,x$;
the call transition on input $a$ pushes $a$ onto the stack, stores the current value of $x$ on the stack,
and resets $x$ to the empty word; and
the return transition on input $b$, while popping the symbol $a$
and stack value $x_p$ from the stack, updates $x$ to $\lt b\, x\, a\rt\,x_p$.

\begin{figure}[htp]
\centering
\input{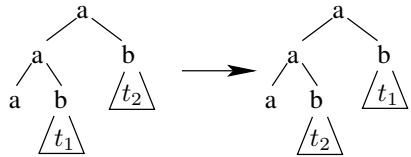}
\caption{Tree Swap}
\label{ex3}
\end{figure}

\mypar{Tree Swap}
Figure~\ref{ex3} shows the transduction that transforms the input tree
by swapping the first (in inorder traversal) $b$-rooted subtree $t_1$
with the next  (in inorder traversal) $b$-rooted subtree $t_2$, not contained in $t_1$,
For clarity of presentation, let us assume that the input word
encodes a tree: it does not contain any internal symbols and
if a call position is labeled $\lt a$ then its matching return is labeled $a\rt$.

The initial state is $q_0$ which means that the transducer has not yet encountered
a $b$-label.
In state $q_0$, the STT records the tree traversed so far using a type-0 variable $x$:
upon an $a$-labeled call, $x$ is stored on the stack, and is reset to $\varepsilon$;
and upon an $a$-labeled return, $x$ is updated to $x_p\lt a\,x\,a\rt$.
In state $q_0$, upon a $b$-labeled call, the STT pushes $q_0$ along with current $x$ on the stack,
resets $x$ to $\varepsilon$, and updates its state to $q'$.
In state $q'$, the STT constructs the first $b$-labeled subtree $t_1$ in variable $x$:
as long as it does not pop stack symbol $q_0$, at a call it pushes $q'$ and $x$,
and at a return, updates $x$ to  $x_p\lt a\,x\,a\rt$ or $x_p\lt b\,x\,b\rt$, depending on
whether the current return symbol is $a$ or $b$.
When it pops $q_0$, it updates $x$ to $\lt b\,x\,b\rt$ (at this point, $x$ contains the tree $t_1$,
and its value will be propagated),
sets another type-1 variable $x'$ to $x_p\,\hole$, and changes its state to $q_1$.
In state $q_1$, the STT is searching for the next $b$-labeled call, and processes $a$-labeled calls and returns
exactly as in state $q_0$, but now using the type-1 variable $x'$.
At a $b$-labeled call, it pushes $q_1$ along with $x'$ on the stack,
resets $x$ to $\varepsilon$, and updates the state to $q'$.
Now in state $q'$, the STT constructs the second $b$-labeled subtree $t_2$ in variable $x$ as before.
When it pops $q_1$, the subtree $t_2$ corresponds to $\lt b\,x\,b\rt$.
The transducer updates $x$ to $x_p'[\lt b\,x\,b\rt]x_p$ capturing
the desired swapping of the two subtrees $t_1$ and $t_2$ (the variable $x'$ is no longer needed
and is reset to $\varepsilon$ to ensure copyless restriction),
and switches to state $q_2$.
In state $q_2$, the remainder of the tree is traversed adding it to $x$.
The output function is defined only for the state $q_2$ and maps $q_2$ to $x$.

\mypar{Tag-based Sorting}
Suppose given a hedge of trees $t_1t_2\cdots t_k$, and a regular pattern, we want to
rearrange the hedge so that all trees that match the pattern appear before the trees that do
not match the pattern. For example, given an address book, where each entry has a tag that denotes
whether the entry is ``private'' or ``public'', we want to sort the address book based on this tag:
all private entries should appear before public entries, while maintaining the original order
for entries with the same tag value.
Such a transformation can be implemented naturally using an STT:
variable $x$ collects entries that match the pattern, while variable $y$ collects entries that do not
match the pattern. As the input is scanned, state is used to determine whether the current tree $t$
satisfies the pattern; a variable $z$ is used to store the current tree,
and once $t$ is read in its entirety, based on whether or not it matches the pattern,
the update $(x,z:=xz,\varepsilon)$ or $(y,z:=yz,\varepsilon)$ is executed.
The output of the transducer is the concatenation $xy$.

\section{Properties and Variants}
In this section, we note some properties and variants of streaming tree transducers
aimed at understanding their expressiveness.
First, STTs compute {\em linearly-bounded\/} outputs, that is,
the length of the output word is within at most a constant factor of the length
of the input word.
The single-use-restriction ensures that
at every step of the execution of the transducer on an input word,
the sum of the sizes of all the variables that contribute to the output term at the end of the execution,
can increase only by an additive constant.

\begin{proposition}[Linear-Bounded Outputs]
For an STT-definable transduction $f$ from $\Sigma$ to $\Gamma$,
for all nested words $w\in\W_0(\Sigma)$, $|f(w)|=\bigO(|w|)$.
\end{proposition}

We now examine some of the features in the definition of STTs in terms of
how they contribute to the expressiveness.
First, having multiple variables is essential, and this follows from results
on streaming string transducers~\cite{AC10,AC11}.
Consider the transduction that rewrites a word $w$ to $w^n$ (that is, $w$ repeated $n$ times).
An STT with $n$  variables can implement this transduction. It is easy to prove
an STT with less than $n$ variables cannot implement this transduction.
Second, the ability to store symbols in the stack at calls is essential.
This is because nested word automata are more expressive than classical finite-state automata over words.

\subsection{Regular Nested-Word Languages}
A streaming tree transducer with empty sets of string variables
can be viewed as an {\em acceptor\/} of nested words:
the input is accepted if the output function
is defined in the terminal state, and rejected otherwise. In this case, the definition coincides
with  (deterministic) nested word automata (NWA).
The original definition of NWAs and regular nested-word languages does not need the input nested
word to be well-matched (that is, the input is a string over $\hat\Sigma$), but this distinction
is not relevant for our purpose.
A nested word automaton $A$ over an input alphabet $\Sigma$ is specified by
a finite set of states $Q$;
a finite set of stack symbols $P$;
an initial state $q_0\in Q$;
a set $F \subseteq Q$ of accepting states;
an internal state-transition function $\delta_i:Q\times \Sigma\mapsto Q$;
a call state-transition function $\delta_c:Q\times \Sigma\mapsto Q\times P$; and
a return state-transition function $\delta_r:Q\times P\times \Sigma\mapsto Q$.
A language $L\subseteq \W_0(\Sigma)$ of nested words is
{\em regular\/} if it is accepted by such an automaton.
This class includes all regular word languages,
regular tree languages, and is a subset of deterministic context-free languages~\cite{AM09}.

Given a nested-word transduction $f$ from input alphabet $\Sigma$ to output alphabet $\Gamma$,
the {\em domain\/} of $f$ is the set $\dom(f)\subseteq \W_0(\Sigma)$ of input nested words $w$
for which $f(w)$ is defined, and the {\em image\/} of $f$ is the set
$\img(f)\subseteq \W_0(\Gamma)$ of output nested words $w'$ such that $w'=f(w)$ for some $w$.
It is easy to establish that for STT-definable transductions, the domain is a regular language,
but the image is not necessarily regular:

\begin{proposition}[Domain-Image Regularity]
\label{prop:img-reg}
For an STT-definable transduction $f$ from $\Sigma$ to $\Gamma$,
$\dom(f)$ is a regular language of nested words over $\Sigma$.
There exists an STT-definable transduction $f$ from $\Sigma$ to $\Gamma$,
such that $\img(f)$ is not a regular language of nested words over $\Gamma$.
\end{proposition}

\subsection{Multi-parameter STTs}

In our basic transducer model, the value of each variable can contain at most one hole.
Now we generalize this definition to allow a value to contain multiple parameters.
Such a definition can be useful in designing an expressive high-level language for
transducers, and will also be used to simplify constructions in later proofs.

We begin by defining nested words with parameters.
The set $\H(\Sigma,\pms)$ of {\em parameterized nested words\/} over the alphabet $\Sigma$
using the parameters in $\pms$,
is defined by the grammar $H := \varepsilon \sep a\sep\pm\sep\lt a\, H\, b\rt\sep H\, H$,
for $a,b\in\Sigma$ and $\pm\in\pms$.
For example, the nested word $\lt a\,\pm_1\,\lt b \rt\,\,\pm_2\, a\rt$ represents
an incomplete tree with $a$-labeled root that has a $b$-labeled leaf as a child
such that trees can be added to its left as well as right by
substituting the parameter symbols $\pm_1$ and $\pm_2$ with nested words.
We can view such a nested word with 2 parameters as a function of arity 2 that
takes two well-matched nested words as inputs and returns a well-matched nested word.

In the generalized transducer model, the variables range
over parameterized nested words over the output alphabet.
Given an alphabet $\Sigma$, a set $X$ of variables, and a set $\pms$ of parameters,
the set $\E(\Sigma,X,\pms)$ of expressions
is defined by the grammar
$E := \varepsilon \sep a\sep\pm\sep x\sep\lt a\, E\, b\rt\sep E\, E\sep E[\pm\mapsto E]$,
for $a,b\in\Sigma$, $x\in X$, and $\pm\in\pms$.
A valuation $\val$ from $X$ to $\H(\Sigma,\pms)$ naturally extends to a function from
the expressions $\E(\Sigma,X,\pms)$ to $\H(\Sigma,\pms)$.

To stay within the class of regular transductions, we need to ensure that
each variable is used only once in the final output and each parameter appears only once in the right-hand side at each
step.
To understand how we enforce single-use-restriction on parameters, consider the update
$x:=xy$ associated with a transition from state $q$ to state $q'$.
To conclude that each parameter can appear at most once in the value of $x$ after the update,
we must know that the sets of parameters occurring in the values of $x$ and $y$ before the update
are disjoint.
To be able to make such an inference statically, we associate, with each state of the transducer,
an occurrence-type that limits, for each variable $x$, the subset of parameters that are allowed to appear in the
valuation for $x$ in that state.
Formally, given parameters $\pms$ and variables $X$,
an {\em occurrence-type\/} $\occ$ is a function from $X$ to $2^{\pms}$.
A valuation $\val$ from $X$ to $\H(\Sigma,\pms)$ is said to be {\em consistent\/} with the occurrence-type $\occ$
if for every parameter $\pm\in\pms$ and variable $x\in X$, if
$\pm\in\occ(x)$ then the parameterized nested word $\val(x)$ contains exactly one occurrence
of the parameter $\pm$, and if $\pm\not\in\occ(x)$ then $\pm$ does not occur in $\val(x)$.
An occurrence-type from $X$ to $\pms$ naturally extends to expressions in $\E(\Sigma,X,\pms)$:
for example, for the expression $e_1 e_2$, if the parameter-sets $\occ(e_1)$ and $\occ(e_2)$ are disjoint,
then $\occ(e_1 e_2)=\occ(e_1)\cup\occ(e_2)$, else the expression $e_1e_2$ is not consistent with the occurrence-type $\occ$.
An occurrence-type $\occ'$ from variables $X$ to $\pms$ is said to be {\em type-consistent\/}
with an occurrence-type $\occ$ from $Y$ to $\pms$ and
an assignment $\rho$ from $Y$ to $X$,
if for every variable $x$ in $X$, the expression $\rho(x)$ is consistent with the occurrence-type $\occ$ and
$\occ(\rho(x))= \occ'(x)$.
Type-consistency ensures that
for every valuation $\val$ from $Y$ to $\H(\Sigma,\pms)$
consistent with $\occ$, the updated valuation $\val\cdot\rho$ from $X$ to $\H(\Sigma,\pms)$
is guaranteed to be consistent with $\occ'$.

Now we can define the transducer model that uses multiple parameters.
A {\em multi-parameter STT\/} $\S$ from input alphabet $\Sigma$ to output alphabet $\Gamma$
consists of states $Q$, initial state $q_0$, stack symbols $P$,
and state-transition functions $\delta_i$, $\delta_c$, and $\delta_r$ as in the case of STTs.
The components corresponding to variables and their updates are specified by
a finite set of typed variables $X$ equipped with a reflexive symmetric binary conflict relation $\eta$;
for each state $q$, an occurrence-type $\occ(q):X\mapsto 2^{\pms}$, and
for each stack symbol $p$, an occurrence-type $\occ(p):X\mapsto 2^{\pms}$;
a partial output function $F : Q \mapsto \E(X,\Gamma,\pms)$ such that
for each state $q$, the expression $F(q)$ is consistent with $\eta$
and $\occ(q)(F(q))$ is the empty set;
for each state $q$ and input symbol $a$, the update function
$\rho_i(q,a)$ from variables $X$ to $X$ over $\Gamma$ is consistent with $\eta$ and it is such that the occurrence-type $\occ(\delta_i(q,a))$
is type-consistent with the occurrence-type $\occ(q)$ and the update $\rho_i(q,a)$;
for each state $q$ and input symbol $a$, the update function
$\rho_c(q,a)$ from variables $X$ to $X$ over $\Gamma$ is consistent with $\eta$ and it is such that,
if $\delta_c(q,a)=(q',p)$ the occurrence-types $\occ(p)$ and $\occ(q')$
are type-consistent with the occurrence-type $\occ(q)$ and the update $\rho_c(q,a)$;
for each state $q$ and input symbol $a$ and stack symbol $p$,, the update function
$\rho_r(q,p,a)$ from variables $X\cup X_p$ to $X$ over $\Gamma$ is
consistent with $\eta$ and it is such that the occurrence-type $\occ(\delta_r(q,p,a))$
is type-consistent with the occurrence-type $\occ(q)$ and $\occ(p)$ and the update $\rho_r(q,p,a)$.

Configurations of a multi-parameter STT are of the form $(q,\sval,\val)$, where $q\in Q$ is a state,
$\val$ is a valuation from variables $X$ to $\H(\Gamma,\pms)$
that is consistent with the occurrence-type $\occ(q)$, and
$\sval$ is a sequence of pairs $(p,\pval)$
such that $p\in P$ is a stack symbol and $\pval$ is a valuation
from variables $X$ to $\H(\Gamma,\pms)$ that is consistent with the occurrence-type $\occ(p)$.
The clauses defining internal, call, and return transitions are as in case of STTs,
and the transduction $\brac{S}$  is defined as before.
In the same as before way we define a copyless multi-parameter STT.

In most of the following proofs we will use the following technique.
We observe that every parallel assignment
can be expressed as a sequence of elementary updates induced by the assignment grammar.
We define this set of elementary updates to be:
1) constant assignment: $x:=w$ where $w$ does not contain variables,
2) concatenation $x:=yz$ where $y$ and $z$ are variables, and
3) parameter substitution: $x:= y[z]$ ($x:= y[\pm\mapsto z]$ for multi-parameters STTs), where
$y$ and $z$ are variables.
In the following proofs we will only consider elementary updates.

Now we establish that multiple parameters do not add to expressiveness:

\begin{theorem}[Multi-parameter STTs]
\label{thm:multi-param}
A nested-word transduction is definable by an STT iff it is definable by a multi-parameter STT.
\end{theorem}

\Proof~
Given an STT $S$ constructing a multi-parameter STT $S'$ is trivial. We use the parameter set $\Pi=\{?\}$, given a state $q$ in $S$,
we will have a corresponding state $q$ in $S'$
and for every type-0 variable $x$ in $q$,
$\occ(q,x)=\emptyset$
while for every type-1 variable $y$ will have $\occ(q,y)=\{?\}$.

We now prove the other direction.
Given a multi-parameter STT
$S=(Q,q_0,P,\Pi,X,\eta,\occ,F,\delta,\rho)$ with $|X|=n$ and $|\Pi|=k$,
we construct an STT $S'=(Q',q_0',P',X',\eta',F',\delta',\rho')$.
We need to simulate the multi-parameter variables using only one hole variables.
We do this by using more hole variables to represent a single multi-parameter variable
and maintaining in the state some information on how to combine them.

The idea is that we maintain a compact representation of every multi-parameter variable.
Consider a variable $x$ with value $\lt a \lt b \pm_1 b\rt \lt c\rt \lt b \pm_2 b\rt a\rt$.
One possible way to represent $x$
using multiple variables, each with only one parameter in its value,
is the following:
$x_1=\lt a ? a\rt$,
$x_2=\lt b ? b\rt \lt c\rt$,
$x_3=\lt b ? b\rt$,
and maintaining in the state the information regarding how to combine
these three values to get $x$. For this, we use a function of the form
$f(x_1)=(x_2,x_3), f(x_2)=\pm_1, f(x_3)=\pm_2$ that tells us to replace the $\hole$ in $x_1$ with $x_2x_3$
and the holes in $x_2,x_3$ with $\pm_1,\pm_2$, respectively. Intuitively
the function $f$ encodes the shape of the tree for every variable in $X$.
The state also needs to remember the root of the tree corresponding to each variable.
We do this with an additional function $g$: $g(x)=x_1$ means that $x_1$ is the root of
the symbolic tree representing $x$.

We now formalize this idea.
$X'$ will contain at most $(2k-1)n$ variables of type-1 and $n$ variables of type-0.
At every step, assuming we are in state $q$,
every variable $x$ will have $2|\occ(x)|-1$ corresponding type-1 variables that represent it
if $\occ(x)\not =\emptyset$,
and one type-0 variable if $\occ(x)=\emptyset$.
Since $\occ(x)\leq k$ at every step we can assume that for every variable $x\in X$,
there are exactly $2k-1$ variables in $S'$ corresponding to it. We denote this set by $V(x)$.

The states in $Q'$ are triplets containing: $q\in Q$, $g:X\mapsto X'$,
$f:X'\mapsto (X'\times X')\cup\Pi\cup\{\varepsilon\}$.
At every step in the computation, each multi-parameter variable in $X$ is represented as a tree over $X'$.
The function $f$ maintains the symbolic
shape of such a tree. The function $g$ tells us, given a variable in $X$
what is the variable in $X'$ representing the root of the tree.
We are going to have $|Q|\cdot (|X'|^2+|\Pi|+1)^{|X'|}\cdot |X'|^{|X|}$ states, where $|X'|=2|\Pi|\cdot |X|$.

There is still a technicality to deal with: at every step,
to maintain the counting argument,
we need to compress the shape $f$ using the observation that
we do not need internal nodes to represent only one parameter.
Whenever this happens, we can just replace the node with its child,
since one variable is enough to represent one parameter. We call this step compression.

We now define the unfolding $f^*$ of the function $f$ that, given a variable in $x\in X'$ provides the corresponding multi-parameter content that it represents:
\begin{itemize}
  \item $f^*(x)=x$ if $f(x)=\varepsilon$
  \item $f^*(x)=x[\pm_i]$  if $f(x)=\pm_i$
  \item $f^*(x)=x[f^*(y)f^*(z)]$ if $f(x)=(y,z)$
\end{itemize}
We then maintain the following invariant at every point in the computation:
the evaluation of $f^*(g(x))$ in $S'$ is exactly the same as the evaluation of $x$ in $S$.

At the beginning every variable is initialized to $\varepsilon$
and so we can represent it with $g(x)=x'$ (where $x'\in X'$ is the type-0 variable corresponding to $x\in X$) and $f(x')=\varepsilon$.
Here the desired invariant about $f^*$ clearly holds.

Let us give the construction at every possible elementary update.
Consider a state $(q,f,g)$ (we only write the parts
that are updated and skip the trivial cases):
\begin{description}
  \item[$\mathbf{x:=w}$:] where $\mathbf{w}$ is a constant
      in the same way as we show in the previous example the content of $x$ can be summarized with $|\occ(x)|$ variables.
  \item[$\mathbf{x:= yz}$:]
      we want to reflect the update in the functions $f$ and $g$.
      First of all we copy the variables in $V(y),V(z)$ into two disjoint subsets in $V(x)$
(we can do this since for the consistency restriction ensures that $\occ(y)+\occ(z)\leq k$.
      We then need to create a new node to be the new root of the two subtrees referring to $y$ and $z$ and update consistently the shapes.
      By induction hypothesis $y$ and $z$ use $2(\occ(y)+\occ(z))-2\leq 2k-2$ variables. So we can still use at least $1$ variable.
      We take unused $x_1'\in V(x)$.
      Then $g'(x)=x_1',f'(x_1')=c(g(y))c(g(z))$ where $c(v)$ is the copy in $V(x)$ of $v$. Compress the result.

  \item[$\mathbf{x:= y{[\pm\mapsto z]}}$:]
      as before we copy the variables in $V(y),V(z)$ into two disjoint subsets in $V(x)$.
      We update the variable containing $\pm$ in $y$ to the tree representing $z$ and we update the corresponding variable.
      Basically after having copied, take $x'$ such that $f(x')=\pm$ and $x'\in V(x)$ belongs to the tree rooted in $c(g(y))$, then
      $f'(x')=f(g(z)),x':=x'[g(z)]$. The counting argument still holds due to the bounded number of parameters.

\end{description}
\begin{figure}
\centering
\input{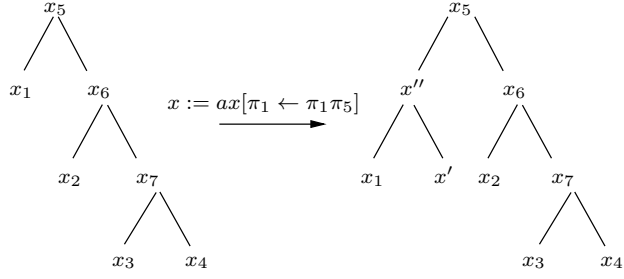}
\caption{Parameter tree for the variable $x=\pm_1\pm_2\pm_3\pm_4$.
In this case (on the left) $g(x)=x_5$ and
$f(x_5)=(x_1,x_6),f(x_6)=(x_2,x_7),f(x_7)=(x_3,x_4),f(x_1)=\pm_1,f(x_2)=\pm_2,f(x_3)=\pm_3,f(x_4)=\pm_4$.
Each variable is of type-1.
After the update we have that $x_5:=ax_5$ and we take two fresh variables $x',x''$ to update the tree
to the one on the right where we set $f(x'')=(x_1,x'),f(x')=\pm_5$.
Since we have 5 parameters and 9 nodes, the counting argument still holds.
Before the update $f^*(x_5)$ evaluates  to $\pm_1\pm_2\pm_3\pm_4$ and after the update
$f^*(x_5)$ evaluates to $a\pm_1\pm_5\pm_2\pm_3\pm_4$.}
\label{paramtree}
\end{figure}
Figure~\ref{paramtree} shows an example of update involving a combination of elementary updates.

We still have to show what happens when we have a call or a return. It's actually easy to see that
the functions at every point can be stored on the stack at a call and recombined at a return with a similar
construction, since all variables are reset at calls.

We need to define the conflict relation $\eta'$ such that the single use restriction is preserved.
For all $x\not =y\in X$ such that $\eta(x, y)$ holds,
then for all $x'\in V(x),y'\in V(y)$, $\eta'(x', y')$ holds.
Also for all $x'\in X'$,  $\eta'(x', x')$ holds.
Since all the assignments that involve conflicting variables are reflected by assignment
over the corresponding trees, this construction is consistent with the new conflict relation.
\qed

\subsection{Bottom-up Transducers}

A nested-word automaton is called {\em bottom-up\/} if it resets its state along the call transition:
if $\delta_c(q,a)=(q',p)$ then $q'=q_0$.
The well-matched
nested word sandwiched between a call and its matching return is
processed by a bottom-up NWA independent of the outer context.
It is known that bottom-up NWAs are as expressive as NWAs over well-matched words~\cite{AM09}.
We show that a similar result holds for transducers also:
there is no loss of expressiveness if the STT is disallowed to propagate information
at a call to the linear successor.
Note than every STT reinitializes all its variables at a call.
An STT $S$ is said to be a {\em bottom-up STT\/} if
for every state $q\in Q$ and symbol $a\in\Sigma$,
if $\delta_c(q,a)=(q',p)$ then $q'=q_0$.

\begin{theorem}[Bottom-up STTs]\label{thm:bottomup}
Every STT-definable transduction is definable by a bottom-up STT.
\end{theorem}

\Proof~
Let $S$ be an STT with states $Q$,
initial state $q_0$, stack symbols $P$,
variables $X$ with a conflict relation $\eta$,
output function $F$,
state-transition functions $\delta_i$, $\delta_c$, and $\delta_r$,
and variable-update functions $\rho_i$, $\rho_c$, and $\rho_r$.
We will construct an equivalent bottom-up STT $S'$.

Given a nested-word $w=a_1a_2\ldots a_k$, for each position $1\le i\le k$,
let $\lwm(w,i)$ be the (well-matched) nested word $a_j,\ldots a_i$, where $j$ is the minimal index
$l$ such that $a_j,\ldots a_i$ is well-matched. Formally
given a well-matched
nested word $w = a_1a_2\ldots a_k$, let us inductively define $\lwm(w,i)$ in the following manner: let
$\lwm(w,0) = \varepsilon$, and for $1 \leq j \leq k$, if position $j$ is internal, then $\lwm(w,j) = w^{j-1}a_j$;
if position $j$ is a call position, then $\lwm(w,j) = \varepsilon$;
and if position $j$ is a return position with the matching call at position $i < j$, then
$\lwm(w,j) = \lwm(w,i-1) a_i \lwm(w,j-1) a_j$.
Each such $\lwm(w,j)$ is well-matched, and represents the subword from the innermost
unmatched call position up to position $j$. For a well-matched word $w$,
$\lwm(w,k)$ equals $w$.
Moreover let $\lc(w,i)$ denote the last unmatched call at position $i$.
If we the first position $j$ in $\lwm(w,i)$ is greater then $1$, then $\lc(w,i)=j-1$ and otherwise it
is undefined.

Since $S'$ must reinitialize its state after a call, at every step,
the state will keep track of the state of $S$ for every possible starting state after the call.
Intuitively, $S'$ delays the application of a call transition of $S$,
and computes the summary of all possible executions of $S$ on the subword between a call and
the corresponding matching return, and this summary can be combined with the information stored
on the stack to continue the simulation after the return.
For this purpose, the state of $S'$ keeps a function $f:Q\mapsto Q$.
When reading the $i$-th symbol of $w$, $f(q)$ represents the state that $S$ would have reached reading
the subword $\lwm(w,i)$ starting in state $q$.
The initial value of $f$ is the identity function $f_0$ that maps each state $q$ to $q$.
On an internal symbol $a$, the function $f$ is updated to $f'$ such that
for each state $q$, $f'(q)=\delta_i(f(q),a)$.
At a call symbol $\lt a$, the current value of $f$ is stored on the stack, along with the symbol $a$,
and $f$ is reset to $f_0$.
Given the current value $f$, to process a return symbol $b \rt$, if the popped value
is $f'$ along with the call symbol $a$, then the updated value $f''(q)$ is defined for each
state $q$ as follows. The value $f'(q)=q_1$ is the relevant state of $S$ before the matching call $\lc(w,i-1)$.
Let $\delta_c(q_1,a)=(q_2,p)$. If $f(q_2)=q_3$, then we know that the transducer $S$ goes from
state $q_2$ to state $q_3$ on the subword sandwiched between the matching call and return $\lwm(w,i-1)$.
Then, the updated state $f''(q)$ should be $\delta_r(q_3,p,b)$.

Now let us explain how $S'$ achieves summarization of variable updates of $S$.
When processing a position $i$
for each variable $x \in X$ and state $q\in Q$, we will have a variable $x_q$ that contains the value
of $x$ assuming $S$ started reading $\lwm(w,i)$ in state $q$.
Initially, and upon every call, for each variable $x$ we have $x_q=?$ or $x_q=\varepsilon$, depending
on its type.
At every input $a$ we perform for each variable $x_q$, the update $\rho(f(q),a,x)$
with each variable $y$ appearing in the right-hand side replaced by $y_q$.

We now need to define a conflict relation $\eta'$ and show that the updates are consistent with this relation.
For all $x,y\in X$, for all $q\not =q'\in Q$, $\eta'(x_q, y_{q'})$ holds; and
for all $q\in Q$, for all $x,y$ such that $\eta(x,y)$ holds, $\eta'(x_q, y_q)$ holds.
Clearly all the right-hand side variables of a single assignment
are variables corresponding to the same state, and so,
if every expression originally is consistent with $\eta$, then every new expression is consistent with $\eta'$.
We now shot that if $\eta'(x, y)$ holds, right-hand side for $x'$ contains $x$,
and right-hand side for updating $y'$ contains $y$, then $\eta(x', y')$ holds.
All variables appearing in right-hand sides for updating variables labeled with the same state, always correspond to the same state.
So the two ways we can have a conflict among two assignments are
either if two variables $x_{q_1},y_{q_1}\in X'$ such that $\eta(x, y)$,
appear in two different assignment to $w_q$ and $z_{q'}$ for some $w,z\in X$,
or $x_{q},y_{q'}\in X'$ such that $q\not =q'$, appear in two different assignment to $w_{q_1}$ and $z_{q_1'}$
for some $w,z\in X$ and $q_1\not =q_1'$.
For the reason of before the second case is trivial since $q_1$ is different from $q_1'$.
In the former case,
we have indeed that either $q=q'$ or they are different.
But in both the cases one of the conflict rules applies so, again we are done.
\qed

\subsection{Regular Look Ahead}

Now we consider an extension of the STT model in which the transducer can make its decisions
based on whether the remaining (well-matched) suffix of the input word
belongs to a regular language of nested words.
Such a test is called {\em regular look ahead}.
A key property of the STT model is the closure under regular look ahead.
Furthermore, in presence of regular-look-ahead, conflict relation can be trivial,
and thus, copyless STTs suffice.

\mypar{Definition of Regular Look Ahead}
Given a nested-word $w=a_1a_2\ldots a_k$, for each position $1\le i\le k$,
let $\lwms(w,i)$ be the (well-matched) nested word $a_i,\ldots a_j$, where $j$ is the maximal index $l$ such that $a_i,\ldots a_l$ is well-matched.
Thus, $\lwms(w,i)$ is the longest well-matched suffix starting at position $i$.
Then, a look-ahead test at step $i$ can test a regular property of the word $\lwms(w,i)$.
Let $L$ be a regular language of nested words,
and let $A$ be a (deterministic) bottom-up NWA for $\rev(L)$ (such an NWA exists, since
regular languages are closed under the reverse operation~\cite{AM09}).
Then, while processing a nested word,
testing whether the  word $\lwms(w,i)$ belongs to $L$ corresponds to
testing whether the state of $A$ after processing
$\rev(\lwms(w,i))$ is an accepting state of $A$.
Since regular languages of nested words are closed under intersection,
the state of a single bottom-up NWA $A$ reading the input word in reverse can be used to test
membership of the well-matched suffix at each step in different languages.
Also note that since $A$ is bottom-up, its state after reading $\rev(\lwms(w,i))$ is same
as its state after reading $\rev(a_i\ldots a_k)$.
This motivates the following formalization.
Let $w=a_1\ldots a_k$ be a nested word over $\Sigma$, and let
$A$ be a bottom-up NWA with states $R$ processing nested words over $\Sigma$.
Given a state $r\in R$, we define the $r$-look-ahead labeling of $w$
to be the nested word $w_r=r_1r_2\ldots r_k$
over the alphabet $R$
such that for each position $1\le j\le k$,
the call/return/internal type of $r_j$ is the same as the type
of $a_j$, and the corresponding symbol is the state of the NWA $A$ after reading $\rev(a_j\ldots a_k)$
starting in state $r$.
Then the {\em $A$-look-ahead labeling of $w$\/},
is the nested word $w_A=w_{r_0}$.
An {\em STT-with-regular-look-ahead\/} consists of a bottom-up NWA $A$ over $\Sigma$ with states $R$, and
an STT $S$ from $R$ to $\Gamma$.
Such a transducer defines a streaming tree transduction from
$\Sigma$ to $\Gamma$: for an input word $w\in\W(\Sigma)$, the output $\brac{S, A}(w)$
is defined to be $\brac{S}(w_A)$.

\mypar{Closure under Regular Look Ahead}
The critical closure property for STTs is captured by
the next theorem which states that regular look-ahead does not add to the expressiveness of
STTs.
This closure property is key to establishing that STTs can compute all MSO-definable
transductions.

\begin{theorem}[Closure under Regular-Look-Ahead]
\label{thm:stt-reg}
The transductions definable by STTs with regular look-ahead are STT-definable.
\end{theorem}
\Proof~
Let $A$ be an NWA with states $R$, initial state $r_0$, stack symbols $P$, and state-transitions functions
$\delta_i'',\delta_r'',\delta_r''$,
and a bottom-up STT
$S=(Q,q_0,P,X,\eta,F,\delta,\rho)$ over $R$.
We construct an STT
$S'=(Q,q_0,P,X',\eta',F',\delta',\rho')$ equivalent to $S$.
The STT $S'$ will be bottom-up.

We use again the definition $\lwm(w,i)$ that we defined in the proof of theorem \ref{thm:bottomup} denoting
Informally given a nested-word $w=a_1a_2\ldots a_k$, for each position $1\le i\le k$,
let $\lwm(w,i)$ be the (well-matched) nested word $a_j,\ldots a_i$, where $j$ is the minimal index
$l$ such that $a_j\ldots a_i$ is well-matched.
This definition will be useful in establishing correctness of our constructions using induction.
One useful observation is that for a well-matched nested word $w$, and an STT $S$, if
$\delta^*((q,\sval, \val), w) =(q',\sval',\val')$,
then $\sval=\sval'$, and in fact this value does not influence the execution of $S$. Hence,
for a well-matched nested word $w$, we can omit the stack, and write
$\delta^*((q, \val), w) =(q',\val')$.

When processing the $i$-th symbol of the input nested word $w$, the transition of the STT $S$ depends on
the state of $A$ after reading the suffix $\lwms(w,i)$.
Since the STT $S'$ cannot determine this value based on the prefix read so far, it needs to simulate $S$ for every
possible choice of $r\in R$.
We will discuss different state components maintained by $S'$ to achieve this goal.
The STT $S'$ keeps in every state a function
$h : R \mapsto R$ and a function
$f : R \mapsto Q$ such that after reading the $i$-th symbol of the input word $w$,
for every state $r$ of $A$,
$h(r)$ gives the state of $A$ when started in state $r$ after reading $reverse(\lwms(w,i))$, and
$f(r)$ gives the state of $S$ after reading $\lwm(w_r,i)$.

In the initial state, $h$ is the identity function that maps each state $r$ to itself, and $f$ is the
constant function that maps each $r$ to the initial state $q_0$.
Suppose the current functions are $f$ and $h$, and the next symbol is an internal symbol $a$.
The updated values
$f'(r)$ and $h'(r)$, for each state $r$, are calculated as follows. Let $r_1 = \delta_i''(r, a)$. This
means that if $A$ starts reading the current subword in reverse in state $r$, it labels the current position
with $r_1$. Then $h'(r)$ should be set to $h(r_1)$. Note that $f(r_1)$ gives the current state of $S$ under the
assumption that $A$ labels the subword so far starting in state $r_1$, and this state is updated using
the transition function of $S$ using the symbol $r_1$:
$f'(r)$ is set to $\delta_i(f(r_1), r_1)$.
At a call symbol $a$, the current values of $f$ and $h$ are stored on the stack, along with the symbol
$a$, and the two functions are reset to their respective initial values.
Suppose the current functions are $f$ and $h$, the next symbol is a return symbol $b$, and the popped
values are functions $f'$ and $h'$ together with call symbol $a$. The updated value $h''(r)$
is computed for each state $r$ as follows. Let $\delta_c''(r, b) = (r_1, p)$.
If $h(r_1) = r_2$, then we know that the NWA $A$ goes
from state $r_1$ to state $r_2$ on the reversed subword sandwiched between the matching call and return.
Then, the state of $A$ before the call (assuming the state after the return is $r$) is
$r_3 = \delta_r''(r_2, p, a)$. Then, the desired $h''(r)$ is $h'(r_3)$ (note: the pushed value $h'$
summarizes the subword before the call). The updated value $f''(r)$ can now be computed by propagating information forwards. The
state $q_1 = f'(r_3)$ gives the state of $S$ before the call assuming the subword upto the call is processed
starting in state $r_3$. Note that the state of $S$ after the call is guaranteed to be its initial state, and
thus, does not depend on the context (this is where we use the fact that $S$ is bottom-up). The
state $q_2 = f(r_2)$ gives the state of $S$ before the return, and this correctly captures the state of $S$
on the subword sandwiched between the call and return. Set $f''(r)$ to $\delta_r'(q_2,(q_1, r_3), r_1)$.

Finally, let us describe how $S'$ keeps track of the variables.
The set of variables is $X'=\{x_r|x\in X,r\in R\}$.
After processing the $i$-th input symbol $x_r$ contains the value of $x$ in $S$ after reading
$\lwm(w_r,i)$.

Let us now define the new conflict relation.
For all $x,y\in X$ such that $\eta(x, y)$ and for all $r\in R$, we have that $\eta'(x_r,y_{r})$
(preserves the conflicts of $S$) and
for all $x,y\in X$ and for all $r_1\not =r_2\in R$, we have that $\eta'(x_{r_1},y_{r_2})$
(at every point only $1$ $r$ is relevant for the final output).
At every step we update all the variables using the states induced by the transition relation $\delta$ of $A$.
While reading the symbol $a$, if $\delta(q',a)=q$ (reading backward), we will update the variables labeled with $q'$ using those of $q$. Notice that two ``sets''  of variables may use the same ``set'' if $\delta(q_1,a)=\delta(q_2,a)=q$.
However only one of these ``set'' will be used when we reach the end of the input.
In fact $F'$ will only use the variables labeled with $r_0$.

We now show that this construction preserves single use restriction.
The proof is very similar to that of bottom-up STTs.
Clearly all the right-hand side variables of a single assignment are taken from the same state and so if there was no conflict relation on a single right hand side, we still have no conflicts on single right-hand sides.
The harder part to show is that
if $x\eta' y$, $x':=fun(x)$ and $y'=fun(y)$ then $x'\eta y'$ holds.
As we said the all the variables on the right-hand sides of variables labeled with the same states always have the same state.
So the two ways we can have a conflict among two assignments are
either if two variables $x_{r_1},y_{r_1}\in X'$ such that $x\eta y$, appear in two different assignments to $w_r$ and $z_{r'}$ for some $w,z\in X$, or $x_{r},y_{r'}\in X'$ such that $r\not =r'$, appear in two different assignment to $w_{r_1}$ and $z_{r_1'}$ for some $w,z\in X$ and $r_1\not =r_1'$.
For the reason of before the second case is trivial since $r_1$ is different from $r_1'$. In the former case we have to do a bit of reasoning. We have indeed that either $r=r'$ or they are different. But in both the cases one of the conflict rules applies so, again we are done.
\qed

\mypar{Copyless STTs with RLA}
Recall that an STT is said to be copyless if $\eta$ only contains the reflexive relation.
In an STT, an assignment of the form $(x,y):=(z,z)$ is allowed if $x$ and $y$ are guaranteed not to be combined,
and thus, if only one of $x$ and $y$ contributes to the final output.
In presence of regular-look-ahead test, the STT can check which variable contribute to the final output,
and avoid redundant updates, and can thus be copyless.

\begin{theorem}[Copyless STT with RLA]
\label{thm:stt-copyless-reg}
A nested-word transduction $f$ is STT-definable iff
it is definable by a copyless STT with regular-look-ahead.
\end{theorem}
\Proof~
One direction is immediate consequence of the closure under RLA of STTs.
We now need to prove the other direction.
Let $S$ be a bottom-up STT with states $Q$,
initial state $q_0$, stack symbols $P$,
variables $X$ with conflict relation $\eta$,
output function $F$,
state-transition functions $\delta_i$, $\delta_c$, and $\delta_r$,
and variable-update functions $\rho_i$, $\rho_c$, and $\rho_r$.
We create a \emph{copyless} STT $S'$ and an RLA automaton $A$
such that $\brac{S', A}(w)$ is equivalent $\brac{S}(w)$.
$S'$ hast states $Q'$,
initial state $q_0'$, stack symbols $P'$,
variables $X$,
output function $F$,
state-transition functions $\delta_i'$, $\delta_c'$, and $\delta_r'$,
and variable-update functions $\rho_i'$, $\rho_c'$, and $\rho_r'$.
$S'$ will not be bottom-up.

We construct a bottom-up automaton $A$ over an alphabet $R$ such that a state
$r\in R$ contains information about which variables will contribute to the final output.
$S'$ will then use the same set of variables of $S$ but at every point it will update only those
contributing to the final output, and resets the others.

Let us first of all prove that if we only update the variables contributing to the final output,
the update function is copyless. This is the same as proving that the set of contributing
variables does not form a conflict.
But this is the definition of conflict relation! This can easily verified using induction with the output function as base case.

Now we show the construction of the automaton $A$.
Since $A$ has to be bottom-up when reading a call (return in the input) we will have to reset the state and this will require some extra bookkeeping.
Particularly after a call we will have to compute the contributing variables without knowing (due to the reset) what's the contributing set of variables at the call.
We now show the construction.

Every state $r$ is going to be a tuple $((s,h_1,h_2),f,g)$
where $s$ is the next symbol of the input string,
$f$ is a partial function from $Q\times 2^X$ to $2^X$,
$g$ is a partial function from $Q$ to $Q$,
$h_1$ is a partial function from $2^X$ to $2^X$, and
$h_2$ is a partial function from $Q\times 2^X$ to $2^X$.
Given the input word $w$, after processing the $i$-th symbol in the input (remember that $A$ reads backward) then
\begin{itemize}
  \item $f(q,Y)=Y'$ if, when $S$ reads $\lwms(w,i)$ starting in state $q$,
    assuming the set of contributing variables at the end of $\lwms(w,i)$ is $Y$, $Y'$ is the current set of relevant variables.
  \item $g(q)=q'$ if, when $S$ reads $\lwms(w,i)$ starting in state $q$, ends in $q'$.
  \item $h_1(q,Y)=Y'$ if, when $S$ reads $\lwms(w,i)$ starting in state $q$,
    assuming the set of contributing variables at the end of $\lwms(w,i)$ is $Y$, and
    $\lc(w,i+1)=j$, then $Y'$ is the set of contributing variables at the end of $\lwms(w,j)$.
  \item $h_2(q,Y)=Y'$ if, when $i$ is call position, and $S$ reads $\lwms(w,i)$ starting in state $q$,
    assuming the set of contributing variables at the end of $\lwms(w,i)$ is $Y$,
    then $Y'$ is the set of contributing variable at position $\ret(w,i)$. $\ret(w,i)$ is the position of
    the return matching the call in position $i$.
\end{itemize}

The information stored in $h_1$ and $h_2$ is useful when processing at call symbols,
and is used by $S'$ to maintain the necessary set of relevant variables
as it explores the hierarchical structure.

The initial state $r_0\in R$ of $A$ will be $((\varepsilon,h_{10},h_{20}),f_0,g_0)$ where $h_{10},h_{20}$ are always undefined, $f_0(q,Y)=Y$ and $g_0(q)=q$.
The initial state $q_0'$ of $S'$ will be $(q_0,\{x_f\})$.

We now show how the state of $A$ is updated.
We assume we are in the state $((s,h_1,h_2),f,g)$ ($h_1$ and $h_2$ are defined only on calls)
and we show how to compute
$((s',h_1',h_2'),f',g')$ (remember that $A$ reads the word backward) on input $s'$.
\begin{description}
  \item[$s'$ is an internal symbol: ]
     in this case $f',g'$ will be simply updated using the transition function of $S$ in the following way:
     $f'(q,Y)=Y'$ where $\delta_i(q,s')=q'$, $f(q',Y)=Y''$ and
     $Y'$ is the union of the variables in the RHS of $\rho_i(q,s',x)$ for all $x\in Y''$ and $g'(q)=g(q')$.

  \item[$s'$ is an return symbol: ]
     processing a return backward, is actually a call for $A$. Since $A$ is bottom up the state that we will reach will always be $r_0$. The current state along with the return symbol is propagated on the stack (that is, $f'=f$, $g'=g$, $h_1'=h_1$ and $h_2'=h_2$).

  \item[$s'$ is an call symbol: ]
     let's call for simplicity $f_p,g_p,h_{1p},h_{2p},s_p$ the components received from the stack at the call (a return reading backward).
     $f'(q_1,Y_1)=Y_2$ and $g'(q_1)=q_4$ and $h_1'(q_1,Y_1)=Y$ and $h_2'(q_1,Y_1)=Y_3$ where,
     $\delta_c(q_1,s')=(q_0,p)$, $g(q_0)=q_2$, $\delta_r(q_2,p,s_p)=q_3$, $g_p(q_3)=q_4$,
     $f_p(q_3,Y_1)=Y_3$
     and $Y_p,Y$ are the set of stack and normal variables on the right hand side of
     $\rho_r(q_2,p,s_p,x)$ for all $x\in Y_3$,
     and $Y_2$ are the variables on the right hand side of
     $\rho_c(q_1,s',x)$ for all $x\in Y_p$.

\end{description}

Now we need to define the update functions for $S'$.
The states $Q'$ of $S'$ are pairs over $Q\times (2^X\cup\{x_f\})$.
After processing the $i$-th position in the input, $S'$ is in state $(q,y)$,
if: 1) $q$ is the state reached by $S$ when processing $\lwm(w,i)$ starting in $q_0$ (since it is bottom-up),
and 2) $Y$ is the set of variables contributing to the final output at the end of $\lwms(w,i+1)$.
We assume without loss of generality that the output in each state is some special assignment to a variable $x_f$.
The stack symbols $P'$ of $S'$ are tuples over $P\times \Sigma \times (2^X\cup\{x_f\}) \times (2^X\cup\{x_f\})$.
The role of the four components will be clear in the construction.

The initial state of $S'$ is $q_0'=(q_0,\{x_f\})$.
Let's assume $S'$ is in state $(q,Y)$ and it reads the input symbol
$a=((s,h_1,h_2),f,g)$.
\begin{description}
  \item[$s$ is an internal symbol: ]
     $\delta_i'((q,Y),a)=(q',Y)$ where $q'=\delta_i(q,s)$ and
     $\rho_i'((q,Y),a,x)=\rho_i(q,s,x)$ if $x\in f(q,Y)$, and $\rho_i'((q,Y),a,x)=\varepsilon$ otherwise.

  \item[$s$ is an call symbol: ]
     $\delta_c'((q,Y),a)=(q_0,Y'),(p,a,Y,Y'')$ where $(q,p)=\delta_c(q,s)$, $Y'=h_1(q,Y)$, $Y''=h_2(q,Y)$,
     $\rho_c'((q,Y),a,x)=\rho_c(q,s,x)$ if $x\in f(q,Y)$, and $\rho_c'((q,Y),a,x)=\varepsilon$ otherwise.

  \item[$s$ is an return symbol: ]
     $\delta_r'((q,Y),(p,a_p,Y',Y''),a)=(q',Y')$ where $q'=\delta_r(q,p,s)$ and \\
     $\rho_r'((q,Y),(p,a_p,Y',Y''),a,x)=\rho_r(q,p,s,x)$ if $x\in Y''$, or $\rho_r'((q,Y),(p,a_p,Y',Y''),a,x)=\varepsilon$ otherwise.

\end{description}
This concludes the proof.
\qed

\subsection{Closure Under Composition}
Now we proceed to show that STTs are closed under sequential composition.
Many of our results rely on this crucial closure property.

\begin{theorem}[Composition Closure]
\label{thm:closure-comp}
Given two STT-definable transductions,
$f_1$ from $\Sigma_1$ to $\Sigma_2$ and $f_2$ from $\Sigma_2$ to $\Sigma_3$,
the composite transduction $f_2\cdot f_1$ from $\Sigma_1$ to $\Sigma_3$ is STT-definable.
\end{theorem}
\Proof~
Using theorems \ref{thm:stt-copyless-reg} and \ref{thm:bottomup}, we consider $S_1$ to be a copyless STT with RLA and $S_2$ to be a bottom-up STT.
We are now given a copyless STT
$S_1=(Q_1,q_{01},P_1,X_1,F_1,\delta_1,\rho_1)$
with RLA automaton $A$ and a bottom-up STT
$S_2=(Q_2,q_{02},P_2,X_2,F_2,\delta_2,\rho_2)$.
We construct a multi-parameter STT $S$ with RLA automaton $A$.
We then use theorem \ref{thm:multi-param} to remove the multi-parameters and
closure under RLA (theorem \ref{thm:stt-reg}) to show that there exists an equivalent STT.


The main idea is that we want to simulate the possible executions of $S_2$ on the output of $S_1$ in a single execution.
We can do this by keeping a summary of $S_2$ in the state and a bigger set of variables.
At every point our transducer has to remember what the output of $S_2$ would be reading the content of the variables in $S_1$ starting in every possible state. A crucial property of the content of the variables in $S_1$ is that
they contain well-matched words. In this way we do not need to collect any stack information for the possible simulations of $S_2$.

Let's show the intuition with an example. Let's say $S$ has only one variable $x$  and $S'$ has only one variable $y$. At some point in the computation on input $a$, $x$ (whose value was $?$) is updated to $ax[?b]$. We would like to reflect this update on $y$ but we do not know in which state we will start processing the value contained in $x$ (assuming this will contribute to the final output), and we still do not know the value that will be stored in the parameter.
The first piece of information we need to track is that of knowing,
for every possible state $q$, which state we will reach in $S_2$ processing the string in $x$ starting reading its content from state $q$.
The problem is actually harder since we have the parameter. But we can extend this idea and keep a function $f$ in the state, that in this particular moment will store $f(q_1,q_2,x)=\delta_2(q_1,a),\delta_2(q_2,b)$ that are the states that we reach reading the contents of $x$ before and after the parameter, assuming we start reading the part before the $\hole$ in $q_1$ and the one after the $\hole$ in $q_2$.
Now we need to know how $y$ gets updated.
Clearly the update of $y$ depends on which state we start reading $x$.
Again we need consider for every pair of states that process the part on the left and on the right of the $\hole$.
We show an example of how we update when reading on the right of the parameter.
Let's assume $\rho_2(q,y,b)=cy$. At this point we do not know what is the previous value of $y$! Fortunately we can fix this by treating the old value of $y$ as a parameter. This tells us that the parameter alphabet will at least contain a parameter for every variable in $X_2$.
We will have then a variable $g(q',q,(x,R),y)$ that is the value of $y$ after reading the value on the right of the $\hole$ in $x$ starting to read the left part in state $q'$ and the right part in state $q$. This variable at the beginning will be simply set to $y'$, a symbolic parameter representing the value of $y$ right after processing the value that will be stored in the $\hole$. We will then perform the updates following the transition relation. So for the case $\rho_2(q,y,b)=cy$
the value of $g(q',q,(x,R),y)$ will now be $cy'$.
Notice, since $S_2$ is bottom up, if there are pending calls, a summary of the part on the left of a parameter $g(q,(x,L),y)$ will simply contain the value of $y$ when reading the last well-matched stretch before the $\hole$ starting in $q_0$. If there aren't pending calls, then it will contain the value of $y$ when reading $(x,L)$ starting in state $q$.

We now give the formal construction.
We denote with $X_{i,j}$ the set of type-j variables in $X_i$.
The states $Q$ of $S=S_2\cdot S_1$ are tuples $(q,f_0,f_{1l},f_{1r})$ where $q\in Q_1$,
$f_{0}:Q_2\times X_{1,0}\mapsto Q_2$,
$f_{1l}:Q_2\times X_{1,1}\mapsto Q_2$ and
$f_{1r}:Q_2\times Q_2 \times X_{1,1}\mapsto Q_2$.
$f_0(q,x)=q'$ when, if $x$ contains $\alpha\in \W_0(\Sigma_2)$, then $\delta_2^*((q,,\varepsilon,\wc),\alpha)=(q',\varepsilon,\wc)$.
$f_{1l}(q_1,x)=q_1'$ when,
if $x$ contains $\alpha\hole\beta\in\W_1(\Sigma_2)$ then $\delta_2^*((q_1,\varepsilon,\wc),\alpha)=(q_1',\wc,\wc)$.
$f_{1r}(q_1,q_2,x)=q_2'$ when,
if $x$ contains $\alpha\hole\beta\in\W_1(\Sigma_2)$ and $\delta_2^*((q_1,\varepsilon,\wc),\alpha)=(\wc,\sval,\wc)$ then
$\delta_2^*((q_2,\sval,\wc),\beta)=(q_2',\varepsilon,\wc)$.

The function $f_0$ (respectively $f_{1l}$ and $f_{1r}$) keeps track of which state $S_2$ would reach reading the content of a variable of type-0 (respectively type-1) of $S_1$
starting in any given state.

We now show how we maintain the invariants defined above at every update.
We assume we are in state $(q,f_0,f_{1l},f_{1r})$ and we only write the parts
that are updated. As before we only consider elementary updates.
We analyze the type-1 case (the 0 case is easier). At every step we indicate with $f_l',f_r',g'$ the updated functions.
\begin{description}
  \item[$\mathbf{x:=w}$:] where $\mathbf{w}$ is a constant $\alpha\hole\beta\in\W_1(\Sigma_2)$.
      Let $(q_1',\sval,\wc)=\delta_2^*((q_1,\varepsilon,\wc),\alpha)$ and
      $(q_2',\varepsilon,\wc)=\delta_2^*((q_2,\sval,\wc),\beta)$ in
      $f_{1l}'(q_1,x)=q_1'$ and
      $f_{1r}'(q_1,q_2,x)=q_2'$.

  \item[$\mathbf{x:= yz}$:]
      we consider without loss of generality the case where $y$ is a type-0 variable and $x,z$ are type-1. We simply use the function stored in the previous state to ``synchronize'' the states of $y$ and $z$.\\
      Let $q_1'=f_{0}(q_1,y)$ in
      $f_{1l}'(q_1,x)=f_{1l}(q_1',z)$ and
      $f_{1r}'(q_1,q_2,x)=f_{1r}(q_1',q_2,z)$.

  \item[$\mathbf{x:= y{[z]}}$:]
      we consider the case where $x,y,z$ are type-1 variables (the other one is simpler).
      We need to ``synchronize'' the left parts and the right parts to update the function $f$.\\
      Let $q_1'=f_{1l}(q_1,y)$ and
      $q_2'=f_{1r}(q_1',q_2,z)$ in
      $f_{1l}'(q_1,x)=f_{1l}(q_1',z)$ and
      $f_{1r}'(q_1,q_2,x)=f_{1r}(q_1,q_2',y)$.

\end{description}

We now define what are the variables $X$ of $S$.
Variables are going to be defined by the union of the following tuples:
$g_0:Q_2\times X_{1,0} \times X_2$,
$g_{1l}:Q_2 \times X_{1,1} \times X_2$ and
$g_{1r}:Q_2 \times Q_2 \times X_{1,1} \times X_2$.
Variable values range over $\Sigma_3\cup \Pi$
where $\Pi=\{x'|x\in X_2\}\cup \{\hole\}$.
$g_0(q_1,x,y)$ is the variable representing the value of $y$ in $S_2$ after reading the content of $x$ (of type-0) of $S_1$
starting in state $q_1\in Q_2$.
The parameters appearing $y$ are symbolic representation of the variable values of
$S_2$ when it starts processing the value stored in the variable $x$.
For example, if $g_0(q_1,x,y)$ contains the value $ay'$, it means that $y'$ is the parameter representing the value of $y$ in the state $q_1$ when
we start reading the content of $x$.
$g_{1l}(q_1,x,y)$ is the value of $y$ after reading the content of $x$ (of type-1)
on the left of the hole assuming we start reading the content of $x$
in state $q_1$. Notice again that, since $S_2$ is bottom up, if there are pending calls, this value will be the same for every $q\in Q_2$ since we only consider the last well-matched stretch in the left part of $x$.
$g_{1r}(q_1,q_2,x,y)$ is the value of $y$ after reading the content of $x$
on the right of the $\hole$ assuming we start reading the left part of $x$
in state $q_1$ and the right part in state $q_2$.

The careful reader will notice that the parameter alphabet also contains $\hole$. Indeed $S_2$ will be using type-1 variable and we still need to
deal with this kind of update. When $\hole$ appears in the $g$ representation of a variable we do not have to worry too much about it and we can treat it as a normal parameter. The problem occurs in the following situation: let's say at a particular step $g(q,x,y)=y'$ but $y$ is a type-1 variable. This can only mean that the $\hole$ appears in $y'$. Now let's assume the next update is of the form $y:=y[a]$. As we can see we still do not have the $\hole$ appearing in the representation of $y$.
We record this fact with a function and delay the substitution using an extra variable for the parameters.
We give an intuition of how to handle this issue but we do not show the full construction for sake of readability.
The next paragraph provides an informal explanation of how to perform symbolic updates and substitution in the summarized variables.

As an example, suppose that at some point the values $x$ and $y$ are $x',y'$ (they both have holes).
We use the variables $x_?=?,y_?=?$ to represent their parameters.
Then, after processing a well-matched subword, we may have an update of this form
$x:= aba x[cc y [a?c]] bb$ and $y:= ab? $.
Notice that the reflexivity of $\eta$ ensures that $x'$ and $y'$ can appear at most once
in the valuation of a variable at any point. This configuration will be captured by
(assuming $q$ is fixed)
$x:=aba x bb$,
$x_? = cc y$,
$y_? = a ? c$ and
$y=ab ?$.
In addition we need to keep information on where the actual parameter of every variable is. We use a function $p:Q\times X \mapsto X^*$ where $p(q,x)$ doesn't contain the same symbol twice for every $x$ and $q$ (this implies boundedness).
The function $p$ will record $p(q,x)=xy$ and $p(q,y)=\varepsilon$. This means that if now we want to reflect the update $x:=x[a]$
we need to perform $x:=x[x'\mapsto x_?[y'\mapsto y_?[?\mapsto a]]]$. In the following we ignore the details regarding the variables of the form $x_?$. Notice that they do not change the form of the construction since are only used as place holder for the summarization.

Now let's come back to our variable summarization.
We now show the updates performed in $S$ for every elementary update in $S'$. We assume we are in a state $q_{cur}=(q,f_0,f_{1l},f_{1r})$ (we only write the parts
that are updated).
We analyze the type-1 cases. We assume the occurrence-type function $\occ:Q\times X\mapsto\Pi$ to be well defined
according to the following assignments (we will prove consistency later).
\begin{description}
  \item[$\mathbf{x:=w}$:] where $\mathbf{w}$ is a constant $\gamma\hole\beta\in\W_1(\Sigma_2)$.
     We simply need to simulate $S_2$ on the content of $x$ taking advantage of the fact that it is well-matched.\\
      Let $\val_r(x)=x'$ for all $x\in S_2$.
      Let $(\wc,\sval,\val_1)=\delta_2^*((q_1,\varepsilon,\val_r),\gamma)$ and
      $(\wc,\varepsilon,\val_2)=\delta_2^*((q_2,\sval,\val_r),\gamma)$.
      Then we have
      $g_{1l}'(q_1,x,y):= \val_1(y)$ and
      $g_{1r}'(q_1,q_2,x,y):= \val_2(y)$.

  \item[$\mathbf{x:= yz}$:]
      we consider without loss of generality the case where $y$ is a type-0 variable and $x,z$ are type-1.
      We need to substitute the values of the variables after reading $y$ in the corresponding parameters in $z$ in order to simulate the concatenation.\\
      Let $q_1'=f_0(q_1,y)$ and $q_1''=f_{1l}(q_1',z)$ in
      $g_{1l}'(q_1,x,u):=g_{1l}(q_1',z,u)[u_i'\mapsto g_{0}(q_1,y,u_i)]$ for all $u_i\in \occ(q_{cur},g_{1l}(q_1',z,u))$ and
      $g_{1r}'(q_1,q_2,x,u):=g_{1r}(q_1',q_2,z,u)$.

  \item[$\mathbf{x:= y{[z]}}$:]
      we consider the case where $x,y,z$ are type-1 variables.
      We need to ``synchronize'' the variables representing the left and right parts in a way similar to the previous case.\\
      Let $q_1'=f_{1l}(q_1,y)$,
      $q_2'=f_{1r}(q_1',q_2,z)$,
      $q_1''=f_{1l}(q_1',z)$,
      $q_2''=f_{1r}(q_1,q_2',y)$ in
      $g_{1l}'(q_1,x,u):=g_{1l}(q_1',z,u)[u_i'\mapsto g_{1l}(q_1,y,u_i)]$ for all $u_i'\in \occ(q_{cur},g_{1l}(q_1',z,u))$ and\\
      $g_{1r}'(q_1,q_2,x,u):=g_{1r}(q_1,q_2',y)[u_i'\mapsto g_{1r}(q_1',q_2,z,u_i)]$ for all $u_i'\in \occ(q_{cur},g_{1r}(q_1,q_2',y))$.

\end{description}

We now need to show that the above construction preserves the single use restriction.
First of all we need to show that the assignments are \emph{consistent} with respect to the parameters.
We actually show a slightly stronger result that will be useful later: the set of variables in $X_2$ corresponding to the parameters appearing in the
right-hand side of a single variable $g(\ldots)$ never violates the conflict relation $\eta_2$ of $X_2$.
Formally, for every $x\in X$, $q\in Q$, $u,v\in\occ(q,x)$, $(u,v)\not\in\eta_2$ and particularly $u\not = v$.
This results is intuitively immediate from the definition of conflict relation. Let's assume by contradiction
that at some point in the computation some parameters $u',v'$ appearing in $\occ(q,x)$ and $u\eta_2 v$.
This means that there exists a run of $S_2$ in which two $u$ and $v$ flow into $x$. But this cannot happen
otherwise we violate the single use restriction.

We now need to show that there exists a conflict relations $\eta$ over the new set of variables consistent with the proposed updates.
We know that the assignments in $S_1$ are copyless.
Thanks to this we know that every time we have an assignment of the form $x:=yz$ or $x:=y[z]$ then $y\not =z$.
Inspecting the updates we perform it is easy to see that, for whatever $\eta$ we will pick the reflexivity will not be violated (the same variable will not appear twice on the same right-hand side).

We now add the following constraints and show that they are consistent with the assignments. For all $q_1,q_1',q_2,q_2'\in Q_2,\ x\in X_{1,0},\ y\in X_{1,1},\ u,v\in X_2$, if $u\eta_2 v$ then
1) $g_0(q_1,x,u)\eta g_0(q_1',x,v)$,
2) $g_{1l}(q_1,y,u)\eta g_{1l}(q_1',y,v)$, and
3) $g_{1r}(q_1,q_2,y,u)\eta g_{1l}(q_1',q_2',y,v)$.
Let's assume there exists an assignment which violates the constraints (we indicate in bold the meta-variables of $S$ and in italic those of $S_2$). There are two possibilities:
\begin{enumerate}
  \item $\mathbf{x}\eta \mathbf{y}$ and they both occur on a right-hand side;
  \item $\mathbf{x}\eta \mathbf{y}$, $\mathbf{x'}:=fun(\mathbf{x})$, $\mathbf{y'}=fun(\mathbf{y})$ but $\mathbf{x'}\eta \mathbf{y'}$ doesn't hold.
\end{enumerate}
We already ruled out the first case when $\mathbf{x}=\mathbf{y}$. When $\mathbf{x}\not =\mathbf{y}$ we want that no assignment violates the above constraints. We check that this is true for the three elementary updates cases. The constant is trivial.
For the case $x:=yz$ we have that two variables can only be in conflict inside the parameter substitution part (the $z$ part) since there is only one summary of $x$.
As we showed before, the parameters in $\occ(q_{cur},g_{1l}(q_1',z,u))$ cannot represent two variables $u,v\in X_2$ such that $u\eta_2 v$ so, this case is ruled out. \

We now need to deal with the second possibility.
Before starting is worthy pointing out that every variable $x\in X_1$ will appear in at most one of the assignments
of $S_2$ due to the copyless restriction.
We want to show that it cannot happen that two variables that are in conflict are assigned to variables that are not in conflict. Let's try to analyze when two variables $\mathbf{x},\mathbf{y}$ assigned to different variables can be in conflict. The first case is that of $\mathbf{x}=\mathbf{y}$. In our settings it means that either 1) $\mathbf{x}=g_{1l}(q_1',z,u)$, 2) $\mathbf{x}=g_0(q_1,y,u_i)$ or 3) $\mathbf{x}=g_{1r}(q_1',q_2,z,u)$. In case 1 we have that $\mathbf{x}'=g_{1l}'(q,x,u)$ and $\mathbf{y'}$ must be $g_{1l}'(q',x,u)$ for some $q\not =q'$, and this means that $\mathbf{x'}\eta \mathbf{y'}$. The same reasoning holds for cases 2 and 3. 
When $\mathbf{x}\not =\mathbf{y}$ one of the cases of the conflict relation $\eta$ defined above must hold. In all cases there are two possibilities: either there was a conflict over 2 different variables in $S_2$ or we are summarizing in two different states.
Let's notice that the conflict must be among two variables of the same kind $g_i(...)\eta g_i(...)$. 
We can then rule out cases 1 and 3 where the conflict is trivially also on the left-hand side. We still have to analyze the case where the conflict is in variables over $g_{0}$. $q_1$ and $y$ are fixed so the only case is where $\mathbf{x}=g_0(q,y,u)$ and $\mathbf{y}=g_0(q,y,v)$ and $u\eta_2 v$. But even in this case the left hand side will be in conflict because it has the same form: $\mathbf{x'}=g_0(q_1,x,u)$ and $\mathbf{y'}=g_0(q_1,x,v)$.
For the case $v:=w[z]$ the argument is very similar.

Again we have to deal with calls and returns, but these as usual can be processed storing more information on the stack (functions $f_0,f_{1l},f_{1r}$ and variables). As in the proof of multi-parameter STT we will store on the stack all the information regarding the variables stored on the stack and at a return we will use them to construct the new values for the state. This can be easily done since at every call the variables get stored on the stack and reset. We can in fact do the same with our variables of $S$. Using an argument similar to the previous one, the assignments will not violate the single use restriction. Notice that the fact that the variables are reset at calls is crucial for this construction.

Our final machine will be a multi-parameter STT with RLA. Since we showed that multi-parameter and RLA are feature that the model can simulate (see theorems \ref{thm:multi-param} and \ref{thm:stt-reg}) we are done.
\qed

\subsection{Restricted Inputs}
A nested word captures both linear and hierarchical structure.
There are two natural classes of nested words: strings
are nested words with only linear structure,
and ranked trees are nested words with only hierarchical structure.
Let us consider how the definition of STT can be simplified when
the input is restricted to these two special cases.

\mypar{Mapping Strings}
Suppose we restrict the inputs to contain only internal symbols,
that is, strings over $\Sigma$.
Then the STT cannot use its stack, and we can assume that
the set $P$ of stack symbols is a singleton set.
This restricted transducer can still map strings
to nested words (or trees) over $\Gamma$ with interesting hierarchical structure,
and hence, is called a {\em string-to-tree\/} transducer.
This leads to the following definition:
a \emph{streaming string-to-tree transducer} (SSTT) $\S$
from input alphabet $\Sigma$ to output alphabet $\Gamma$ consists of
a finite set of states $Q$;
an initial state $q_0\in Q$;
a finite set of typed variables $X$;
a partial output function $F : Q \mapsto \E_0(X,\Gamma)$ such that
for each state $q$, a variable $x$ appears at most once in $F(q)$;
a state-transition function $\delta_i:Q\times \Sigma\mapsto Q$;
and a  variable-update function $\rho_i:Q\times \Sigma\mapsto \A(X,X,\eta,\Gamma)$.
Configurations of such a transducer are of the form $(q,\val)$, where $q\in Q$ is a state,
and $\val$ is a type-consistent valuation for the variables $X$.
The semantics $\brac{S}$ of such a transducer is a partial function from
$\Sigma^*$ to $\W_0(\Gamma)$. We notice that in this setting the
copyless restriction is enough to capture MSO completeness
since the model is closed under RLA (i.e. a reflexive $\eta$ is enough).

\begin{theorem}[Copyless String-To-Tree STT are Closed under RLA]
\label{thm:sstt-reg}
The transductions definable by copyless SSTTs with regular look-ahead are also definable by copyless SSTTs.
\end{theorem}
\Proof~
Let $A$ be an DFA with states $R$, initial state $r_0$, and state-transitions function
$\delta_A$ over an input alphabet $\Sigma$.
Given a copyless string-to-tree STT
$S=(Q,q_0,X,F,\delta,\rho)$ over $R$,
we construct an equivalent copyless STT
$S'=(Q',q_0',Z,F',\delta',\rho')$ over $\Sigma$.
Clearly since the input is a string we can consider $S$ to only have transitions of the
form $\delta_i$ and variable updates of the form $\rho_i$ that for simplicity we will denote by $\delta$ and $\rho$.

The transition of the STT $S$ at a given step depends on the state of $A$ after reading
the reverse of the suffix.
Since the STT $S'$ cannot determine this value based on the prefix, it needs
to simulate $S$ for every possible choice.
For a string $w=a_1\ldots a_k$, and a state $r\in R$,
define the string $w_{r}$ over $R$ to be
equal to $r_1r_2\ldots r_k$
such that for each position $1\le j\le k$,
the corresponding symbol is the state of the DFA $A$ after reading $\rev(a_j\ldots a_k)$
{\em starting in state $r$}. At the end of the string we will be interested in $w_A=w_{r_0}$.

We will discuss different state components maintained by $S'$ in $Q'$.
Every state in $Q'$ contains a function $h:R\mapsto R$ and a state $f:R \mapsto Q$
such that after reading the $j$-th symbol,
for every state $r$ of $A$, $h(r)$ gives the state of $A$ when started in state $r$ after
reading $\rev(w_1\ldots w_j)$, and  $f(r)$  gives the state of $S$ after reading $(w_1\ldots w_j)_r$.
The state will also contain two functions $g$ and $p$ that we will discuss later.
In the initial state $q_0'$, $h$ is the identity function that maps each state $r$ to itself,
and $f$ is the constant function that maps each $r$ to the initial state $q_0$.

Suppose the current functions are $f$ and $h$, and the next symbol is a symbol $a$.
The updated values $f'(r)$ and $h'(r)$, for each state $r$, are calculated as follows.
Let $r_1=\delta_A(r,a)$. This means that if $A$ starts reading the current subword in reverse in
state $r$, it labels the current position with $r_1$.
Then $h'(r)$ should be set to $h(r_1)$.
Note that $f(r_1)$ gives the current state of $S$
under the assumption that $A$ labels the subword so far starting in state $r_1$, and this state is updated
using the transition function of $S$ using the symbol $r_1$: $f'(r)$ is set
to $\delta(f(r_1),r_1)$.

Finally, let us describe how $S'$ keeps track of the variables.
For each state $r$ of $A$ and variable $x$ of $S$, $S'$ keeps a copy $g(r,x)$
that is supposed to capture the value of $x$ assuming the next symbol is labeled with $r$.
Let us see how these values can be updated when processing a symbol $a$.
If $r_1=\delta_A'(r,a)$ then the updated values $g'(r,-)$ are obtained from
$g(r_1,-)$ by applying the variable-update function $\rho(f(r_1),r_1)$.
The problem is that there may be another state $r'$ with $r_1=\delta_i'(r',a)$,
and this implies that the updated values  $g'(r',-)$ also depend on $g(r_1,-)$.
This sharing poses a challenge since the update in $S'$ needs to be copyless.
Starting with $g(r,x)=g(r',x)$, the variable-update assignments
can add output symbols at the two ends of this string in different manners
for $g(r,x)$ and $g(r',x)$.

Our solution relies on a symbolic representation and a careful analysis of sharing.
First of all we will need multi-parameters STTs to be able to represent variables. We will create a copyless
multi-parameter STT and then use the fact that the translation from multi-parameter STT to STTs preserves the copyless property.
The STT $S'$ uses a set $Z$ of variables which store actual values of output strings,
and a ``shape'' function $g:R\times X \mapsto T(Z)$ (where be is the set of ordered trees over $Z$, even though
we will only need trees where every variable appears at most once).
The number of variables in $Z$ that we need will be explained shortly.
Given a valuation of all the $z$-variables, we can substitute these values in $g$ to get
the value for each variable of $S$ for a given state-label $r$.

What we really collect in the shape function $g(r,x)=z(z_1,z_2)$, for example, is a way to use the variables to get the current valuation of $x$
assuming the next symbol is $r$. For example the current value of $x$ assuming the next symbol is $r$ in this case is $z[\pi_1 \mapsto z_1,\pi_2\mapsto z_2]$. This example shows that we need multi-parameter STTs.
In particular our parameter alphabet will be $\Pi=\{?\}\cup \{\pi_1,\ldots,\pi_{|X|}\}$ where $?$ is the actual parameter of the variable we are representing and $\{\pi_i\}$ is the parameter representing the i-th children
of a node in the tree.
In our case, if a variable contains a parameter $\pi_i$, it contains also $\pi_j$ for every $j<i$ and they appear in order.

We can immediately se that the parameter $?$ will appear in some position of the tree (we will force it to be a leaf) that may change during the computation.
This tells us that we need a new function in the state recording the position of $?$. We use a function $p:R\times X\mapsto Z\cup\{\varepsilon\}$ that tells us which variable in the tree contains the $?$. $p(r,x)$ is $\varepsilon$ when $x$ is of type-0.
$p(r,x)=z$ means that the variable $z$ contains a value of the form $\alpha ? \beta$.


A tree $t$ over $Z$ is said to be repetition-free if no symbol occurs twice in $t$.
Given two repetition-free trees $t$ and $t'$, a tree $s$ is a maximal shared prefix-subtree
between $t$ and $t'$ if
(1) if there exists two extensions of $s$, $s_1,s_2$ that are subtrees of $t$ and  $t'$ (a tree $s'$ is an extension of a tree $s$
if they can be made equal by deleting zero or more subtrees from $s'$),
(2) $s$ is not a proper subtree of any $s'$ ($s$ is a subtree of $s'$ but they are not the same), such that $s'$ is a shared prefix-tree of both  $t$ and  $t'$, and
(3) $s$ contains at least one node.
Given two repetition-free trees $t$ and $t'$, let $N(t,t')$ denote the number of maximal
shared prefix-subtree between $t$ and $t'$.

Our representation maintains the following invariants for the shapes:
\begin{enumerate}
\item
Each shape $g(r,x)$ is repetition-free.
\item
For all states $r,r'$,
$\sum_{x,y\in X}N(g(r,x),g(r',y))$ is at most $|X|$.
\item
The shapes are compressed: if the subtree $z_1(\ldots,z_2(\ldots),\ldots)$ occurs in a shape $g(r,x)$
and $z_2$ is not equal to $p(r,x)$, then
there must be a shape $g(r',y)$ which either contains $z_1$ but not a subtree of the form $z_1(\ldots,z_2(\ldots),\ldots)$ or
contains $z_2$ but not a subtree of the form $z_1(\ldots,z_2(\ldots),\ldots)$.
In the case where $z_2$ is equal to $p(r,x)$ we require that $z_1$ has more than one child.
\item
$?$s are not shared: for all $r,x$, if $p(r,x)=z$, then $z$ is not a shared and it is a leaf in $g(r,x)$.
\item
No shape contains more than $|\Pi|+1$ leaves.
\end{enumerate}


The first invariant ensure the bounded size of shapes.
Notice that the second invariant implies that
for $x\not=y$, for each $r$, $g(r,x)$ and $g(r,y)$ are disjoint.
The second invariant implies that
for every state $r$, the tree $g(r,x)$, for all $x$ cumulatively, can have a
total of $|X||R|$ maximal shared prefix-subtree with respect to all other strings.
The compression assured by the third invariant then implies that
the sum $\sum_{x\in X}|g(r,x)|$ is bounded by $|X|+2|X||R|$.
As a result it suffices to have $|R|(|X|+2|X||R|)$ variables in $Z$.
The fourth invariant helps us dealing with variable substitution.
Notice that this invariants implies that the variables $p(r,x)$ never contain
any parameter other than $?$.
The fifth invariant guarantees the well formedness of our assignments.

Given a shape $g$ with parameter function $p$, and an internal symbol $a$, to compute the updated values $g'(r,x)$ and $p'(r,x)$, we need to consider the right-hand side $\rho(f(r_1),r_1)(x)$, for $r_1=\delta_A(r,a)$,
and replace each variable $y$ with the current shape $g(r_1,y)$.
As in case of the proof of the lemma, we split the update into a sequence of simpler updates.

Given a shape $g(r,x)$ we denote with $c(r,x,z')$ the sequence of children of the subtree $t$ of $g(r,x)$
such that the root of $t$ is $z'$.
In the following, whenever not stated, we assume that at end of every update a ``normalization'' is applied to avoid violation of the invariants 3 and 4.
For the third invariant this means that, if after an update we have a shape $g(r,x)$ with a subtree $z(z_1\ldots,z_{i-1},z_i,z_{i+1},\ldots z_n)$, such
that both $z$ and $z_i$ occur only in this shape we normalize the shape in the following way:
1) $z$ is set to $z[\pi_i\mapsto z_i]$ and
2) $g(r,x)$ is updated to $z(z_1\ldots,z_{i-1},c(r,x,z_i),z_{i+1},\ldots z_{n})$ and the parameters
in $z$ are consistently renamed. Similarly for the case where $z_i$ contains $?$ and $i=n=1$.

For what concern the fourth invariant, the normalization works as follows.
If after an update we have a shape $g(r,x)$ such that $p(r,x)=z$ is a shared variable (by construction it can
only be a leaf), we do the following:
1) for every $r',x'$ such that $p(r',x')=z$, generate a fresh variable $z_{r',x'}$ and set it to $?$,
2) update $z$ to $z[?\mapsto \pi_1]$
3) in every shape $g(r',x')$ where $z$ appears update the subtree rooted in $z$ inserting $z_{r',x'}$
as the only child and set $p(r',x')$ to $z_{r',x'}$.

We analyze a richer set to elementary updates to better understand the construction.

Consider the case $x:=\langle a x b\rangle$.
If $g(r_1,x)$ has root $z$.
If $z$ does not occur in any other $g(r',y)$, then
we update $g'(r,x)$ to $g(r_1,x)$, and update $z$ to $\langle a z b\rangle$.
If $z$ does occur in some other $g(r',y)$, then we use a ``fresh'' symbol $z_f$ that does not occur
in any $g(\wc,\wc)$, and update the shape $g'(r,x)$ to $z_f(g(r_1,x))$, and set $z_f$ to $\langle a \pi_1 b\rangle$.
Assuming $g$ satisfies the three shape invariants, it is easy to show that the updated shape
continues to satisfy the invariants.
The case of appending a symbol to $x$ is similar.

Consider the case $(x,y):=(y,x)$.
We swap the values of $g(r,x)$ and $g(r,y)$, and this clearly maintains
all the invariants. The reset to $\varepsilon$ case is also trivial.

Consider the assignment $(x,y):=(xy,\varepsilon)$ (where without loss of generality $x$ is of type-1 and $y$ is
type-0).
Suppose $g(r_1,x)$ has root $z_x$, and  $g(r_1,y)$ has root $z_y$.
We have four possible cases in which we update $g(r,x),g(r,y)$ in different ways:
\begin{itemize}
\item none of $z_x$ and $z_y$ occurs in some other $g(r',x')$.
	     In this case we set $z_x$ to $z_xz_y$ and we update $g'(r,x)$ to $z_x(c(r,x,z_x),c(r,y,z_y'))$.
         Doing this we also have consistently renumber the parameters in $z_x$ (we will ignore this detail from
         now on).
         We now have $z_y$ unused so we can assign it to $g'(r,y)$ and update
         it to $\varepsilon$.
		 We do actually have to be careful. In fact we want to preserve the invariant
		 that for all $r',x'$, $p(r',x')$ is a leaf (for the sharing invariant we will normalize later).
		 In this case nothing bad can happen since $y$ is a type-0 variable, but if it was of type
		 1 we would have had to consider the case where $p(r_1,y)$ was equal to $x_y$ and avoid
		 to merge it with other variables. This case is really similar to the case when both $z_x$ and $z_y$
		 are shared.

\item $z_x$ occurs in some other $g(r',x')$ while $z_y$ does not.
         We can't update $z_x$ since it would also change its value in its other occurrences.
         We need therefore to remember the update in the shape.
	     We set $z_y$ to $\pi_1 z_y$ (where $z_y$ parameter are shifted by $1$) and
         we update $g'(r,x)$ to $z_y(g(r_1,x),c(r,y,z_y))$.
         Since the assignment does not violate the third invariant we can take a fresh variable
         $z_f$ that we use to represent the value of $y$ and we update it to $\varepsilon$.
         We then assign $g'(r,y)$ to $z_f'$.
\item $z_y$ occurs in some other $g(r',x')$ while $z_x$ does not. Similar to previous case.
\item both $z_x$ and $z_y$ occur in some other $g(r',x')$, $g(r'',y')$ respectively.
	     Even this case is similar.
         We take a fresh $z_f$ and update it to $z_xz_y$.
         We consistently update $g'(r,x)$ to $z_f(g(r_1,x),g(r_1,y))$.
         The third invariant clearly holds so we can take a free variable to update $g'(r,y)$.
\end{itemize}

Now let's consider the assignment $(x,y):=(x[y],?)$ (where without loss of generality $x$ and $y$ are of type-1.
The updates are going to be similar to those of the previous case. However we need to use the function $p$ to understand how to plug the shapes together).
Suppose $p(r_1,x)$ is $z_{px}$, $g(r_1,y)$ has root $z_y$ and $p(r_1,x)$ is $z_{py}$,.

We have four possible cases in which we update $g(r,x),g(r,y)$ in different ways:
\begin{itemize}
\item none of $z_{px}$ and $z_y$ occurs in some other $g(r',x')$.
	     In this case we set $z_{px}$ to $z_{px[?\mapsto z_y]}$ and we update
         $g'(r,x)$ to $g(r_1,x)$ where we replace the subtree rooted in $z_{px}$ with $g(r_1,y)$.
         We now have that $z_y$ is unused so we can assign it to $g'(r,y)$ and update it to $\varepsilon$.
         To record the position of the parameter we update
         $p'(r,x)$ to $p(r_1,y)$ if different from $z_y$ and we leave unchanged otherwise.
         $p'(r,y)$ is set to $(z_y,0)$.

\item $z_{px}$ occurs in some other $g(r',x')$ or $z_{py}$ does.
         This case violates the fourth invariant, so it can't occur.

\item $z_y$ occurs in some other $g(r',x')$ while $z_{px},z$ and $z_{py}$ do not.
	     In this case we set $z_{px}$ to $z_{px[?\mapsto \pi_1]}$ and we update
         $g'(r,x)$ to $g(r_1,x)$ where we replace the subtree rooted in $z_{px}$ with $z_{px}(g(r_1,y))$.
         Now if the third invariant is violated we can compress $z_{px}$ we can apply the
         normalization and get a free variable $z_f$ otherwise we have it already.
         To record the position of the parameter we update
         $p'(r,x)$ to $r_1,y$ if different from $(z_f,0)$ and we leave unchanged otherwise.
         $p'(r,y)$ is set to $(z_f,0)$.
\end{itemize}

\qed

\mypar{Mapping Ranked Trees}
In a ranked tree, each symbol $a$ has a fixed arity $k$, and an $a$-labeled node
has exactly $k$ children. Ranked trees can encode terms, and existing literature
on tree transducers focuses primarily on ranked trees.
Ranked trees can be encoded as nested words of a special form, and the definition of an STT can
be simplified to use this structure.
For simplicity of notation, we assume that there is a single $0$-ary symbol $\zero\not\in\Sigma$,
and every symbol in $\Sigma$ is binary.
The set $\BT(\Sigma)$ of {\em binary trees\/} over the alphabet $\Sigma$ is then
a subset of nested words defined by the grammar
$T\ :=\ \zero \sep \lt a\, T\, T\, a\rt$, for $a\in\Sigma$.
We will use the more familiar tree notation
$a\lt t_l\, t_r\rt$, instead of $\lt a\,t_l\,t_r\, a \rt$, to
denote a binary tree with $a$-labeled root and subtrees $t_l$ and $t_r$ as children.
The definition of an STT can be simplified in the following way if we know
that the input is a binary tree.
First, we do not need to worry about processing of internal symbols.
Second, we restrict to bottom-up  STTs due to their
similarity to bottom-up tree transducers,
where the transducer returns, along with the state,
values for variables ranging over output nested words,
as a result of processing a subtree.
Finally, at a call, we know that there are exactly two subtrees, and hence, the
propagation of information across matching calls and returns using a stack can be
combined into a unified combinator:
the transition function computes the result corresponding to a tree $a\lt  t_l\,t_r\rt$
based on the symbol $a$, and the results of processing the subtrees $t_l$ and $t_r$.

A \emph{bottom-up ranked-tree transducer} (BRTT) $\S$
from binary trees over $\Sigma$ to nested words over $\Gamma$ consists of
a finite set of states $Q$;
an initial state $q_0\in Q$;
a finite set of typed variables $X$ equipped with a conflict relation $\eta$;
a partial output function $F : Q \mapsto \E_0(X,\Gamma)$ such that
for each state $q$, the expression $F(q)$ is consistent with $\eta$;
a state-combinator function $\delta:Q\times Q \times \Sigma\mapsto Q$;
and a variable-combinator function $\rho:Q\times Q\times \Sigma\mapsto \A(X_l\cup X_r,X,\eta,\Gamma)$,
where $X_l$ denotes the set of variables $\{x_l\sep x\in X\}$,
$X_r$ denotes the set of variables $\{x_r\sep x\in X\}$, and conflict relation $\eta$ extends to these sets naturally.
The state-combinator extends to trees in $\BT(\Sigma)$:
$\delta^*(\zero)=q_0$ and $\delta^*(a\lt  t_l\,t_r\rt)=\delta(\delta^*(t_l),\delta^*(t_r),a)$.
The variable-combinator is used to map trees to valuations for $X$:
$\val^*(\zero)=\val_0$, where $\val_0$ maps each type-0 variable to $\varepsilon$ and each type-1 variable to $\hole$,
and  $\val^*(a\lt  t_l\,t_r \rt)=\rho(\delta^*(t_l),\delta^*(t_r),a)[X_l\mapsto\val^*(t_l)][X_r\mapsto\val^*(t_r)]$.
That is, to obtain the result
of processing the tree $t$ with $a$-labeled root and subtrees $t_l$ and $t_r$,
consider the states $q_l=\delta^*(t_l)$ and $q_r=\delta^*(t_r)$,
and valuations $\val_l=\val^*(t_l)$ and $\val_r=\val^*(t_r)$,
obtained by processing the subtrees $t_l$ and $t_r$.
The state corresponding to $t$ is given by the state-combinator $\delta(q_l,q_r,a)$.
The value $\val^*(x)$ of a variable $x$ corresponding to $t$ is obtained from
the right-hand side $\rho(q_l,q_r,a)(x)$ by setting variables in $X_l$ to values given by
$\val_l$ and setting variables in $X_r$ to values given by $\val_r$.
Note that the consistency with conflict relation ensures that each value gets used only once.
Given a tree $t\in \BT(\Sigma)$, let $\delta^*(t)$ be $q$ and let $\val^*(t)$ be $\val$.
Then, if $F(q)$ is undefined then
$\brac{S}(t)$ is undefined, else $\brac{S}(t)$ equals $\val(F(q))$ obtained by evaluating the
expression $F(q)$ according to valuation $\val$.

\begin{theorem}[Expressiveness of Ranked Tree Transducers]
A partial function from $\BT(\Sigma)$ to $\W_0(\Gamma)$ is STT-definable iff it is BRTT-definable.
\end{theorem}
\Proof~
We give a sketch for the constructions.
We first show that given a BRTT $S$ from $\BT(\Sigma)$ to $\W_0(\Gamma)$
we can construct a STT $S'$. This translation is quite easy. At every call
we store on the state the current information.
Since the input is a binary tree we only need $|X|$ variables, where $X$ is the set of variables of $S$. Let's assume we are in the state $i$ right after a call. Now all the variables are reset and we can process the left child. After that we store the computation on the stack of the right child. At its return we will have the values of $X_l$ on the stack and those of $X_r$ in the variables so we can combine them. Now that we read the matching return of $i$ we can continue the computation in the same way.

We know from theorem \ref{thm:bottomup} that bottom-up STTs are as expressive as STTs.
Given a bottom-up STT $S$ we construct a BRTT $S'$.
Again we know the trees are binary and since $S$ is bottom-up it resets its computation at every call. We omit the details of the proof but we give some intuition. Since the tree is ranked there will not be internal symbol. 
It should be easy to identify, by inspection of the STT rules, the set of leaves and the corresponding computation.
This gives us the first rules in $S'$.
Now we need to construct the internal nodes rules. This can be done in a similar way to that for leaves. We only need to inspect all the return rules and use the state popped from the stack for the computation regarding first child and the current state for the one regarding the second child.
\qed

\subsection{Restricted Outputs}
Let us now consider how the transducer model can be simplified when
the output is restricted to the special cases of strings and ranked trees.
The desired restrictions correspond to limiting the set of allowed operations
in expressions used for updating variables.

\mypar{Mapping  Nested Words to Strings}
Each variable of an STT stores a potential output fragment.
These fragments get updated by addition of outputs symbols,
concatenation, and insertion of a nested word in place of the hole.
If we disallow the substitution operation, then the STT cannot manipulate
the hierarchical structure in the output.
More specifically, if all variables of an STT
are type-0 variables, then the STT produces outputs that are strings
over $\Gamma$.
The set of expressions used in the right-hand sides can be simplified to
$E_0:=\varepsilon \sep a \sep x_0\sep E_0E_0$.
That is, each right-hand side is a string over $\Gamma\cup X$.
Such a restricted form of STT is called a {\em streaming tree-to-string transducer\/} (STST).
While less expressive than STTs, this class is adequate to compute
all tree-to-string transformations, that is, if the final output of an STT
is a string over $\Gamma$, then it does not need to use holes and substitution:

\begin{theorem}[STST Expressiveness]
\label{thm:STST-expressiveness}
A partial function from $\W_0(\Sigma)$ to $\Gamma^*$ is STT-definable iff it is STST-definable.
\end{theorem}

If we want to compute string-to-string transformations, then
the STT does not need a stack and does not need type-1 variables.
Such a transducer is both an SSTT and an STST, and this restricted class
coincides with the definition of streaming string transducers (SST)~\cite{AC11}.

\mypar{Mapping Nested Words to Ranked Trees}
Suppose we are interested in outputs that are binary trees in $\BT(\Gamma)$.
Then, variables of the transducer can take values that range over such
binary trees, possibly with a hole.
The internal symbols, and the concatenation operation, are no longer needed
in the set of expressions.
More specifically, the grammar for the type-0 and type-1 expressions can be
modified as:
\begin{eqnarray*}
E_0 & := & \zero \sep  x_0 \sep a\lt \, E_0\, E_0\,\rt  \sep E_1[E_0]\\
E_1 & := & \hole\sep x_1 \sep a\lt \, E_0\,E_1\,\rt \sep  a\lt \, E_1\,E_0\,\rt  \sep E_1[E_1],
\end{eqnarray*}
where $a\in\Gamma$, $x_0\in X_0$ and $x_1\in X_1$.
To define transformations from ranked trees to ranked trees,
we can use the model of bottom-up ranked-tree transducers with the above grammar.

\section{Expressiveness}

The goal of this section is to prove that the class of nested-word transductions
definable by STTs coincides with the class of transductions definable
using Monadic Second Order logic (MSO).
Our proof relies on the known equivalence between MSO and Macro Tree Transducers
over {\em ranked trees\/}.

\subsection{MSO for Nested Word Transductions}
Formulas in monadic second-order logic (MSO) can be used to define
functions from (labeled) graphs to graphs~\cite{Cou94}.
We adapt this general definition for our purpose of defining
transductions over nested words.
A nested word $w=a_1\ldots a_k$ over $\Sigma$
is viewed as an edge-labeled graph $G_w$ with $k+1$ nodes $v_0\ldots v_k$
such that (1) there is a (linear) edge from each $v_{j-1}$ to $v_j$, for $1\le j\le k$,
labeled with the symbol $a_j\in\Sigma$, and
(2) for every pair of matching call-return positions $i$ and $j$, there is an unlabeled
(nesting) edge from $v_{i-1}$ to $v_{j-1}$.
The monadic second-order logic of nested words is given by the syntax:
\[\phi  :=  a(x,y) \sep X(x) \sep  x \matchrel y \sep \phi\vee\phi \sep
\neg\phi \sep \exists x.\phi \sep \exists X.\phi\]
where $a\in \Sigma$, $x, y$ are first-order variables, and $X$ is a second-order variable.
The semantics is defined over nested words in a natural way.
The first-order variables are interpreted over nodes in $G_w$, while set variables are
interpreted over sets of nodes. The formula $a(x,y)$ holds if there an $a$-labeled edge from
the node $x$ to node $y$ (this can happen only when $y$ is
interpreted as the linear successor position of $x$),
and $x\matchrel y$ holds if the nodes $x$ and $y$ are connected by a nesting edge.

An \emph{MSO nested-word transducer} $\Phi$ from input
alphabet $\Sigma$ to output alphabet $\Gamma$ consists of
a finite copy set $C$,
node formulas $\phi^c$, for each $c\in C$, each of which is an MSO formula over nested words over $\Sigma$ with one free
first-order variable $x$,  and
edge formulas $\phi^{c,d}$ and $\phi^{c,d}_a$, for each $a\in\Gamma$ and $c,d\in C$,
each of which is an MSO formula over nested words over $\Sigma$ with two free first-order variables $x$ and
$y$. Given an input nested word $w$, consider the following output graph: for each node
$x$ in $G_w$ and $c\in C$, there is a node $x^c$ in the output if the formula $\phi^c$ holds
over $G_w$, and for all such nodes $x^c$ and $y^d$, there is an $a$-labeled edge from $x^c$
to $y^d$ if the formula $\phi^{c,d}_a$ holds over $G_w$,
and there is a nesting edge from $x^c$
to $y^d$ if the formula $\phi^{c,d}$ holds over $G_w$.
If this graph is the graph corresponding to the nested word $u$ over $\Gamma$ then $\brac{\Phi}(w)=u$,
and otherwise $\brac{\Phi}(w)$ is undefined.
A nested word transduction $f$ from input alphabet $\Sigma$ to output alphabet $\Gamma$
is {\em MSO-definable\/} if there exists an MSO nested-word transducer  $\Phi$ such that $\brac{\Phi}=f$.

By adapting the simulation of string transducers by MSO~\cite{EH01,AC10},
we show that the computation of an STT can be encoded by MSO, and thus,
every transduction computable by an STT is MSO definable.

\begin{theorem}[STT-to-MSO]
\label{thm:stt-mso}
Every STT-definable nested-word transduction  is MSO-definable.
\end{theorem}
\Proof~
Consider a copyless STT $S$ with RLA automaton $A$.
The labeling of positions of the input word with states of the RLA automaton can be
expressed in MSO.
The unique sequence of states and stack symbols
at every step  of the execution of the transducer $S$ over a given input nested word $w$ can
be captured in MSO using second order existential quantification.
Thus, we assume that each node in the
input graph is labeled with the corresponding state of the STT while processing the next symbol.
The positions corresponding to calls and returns are additionally labeled with
the corresponding stack symbol pushed/popped.

\begin{figure}[htp]
\centering
\input{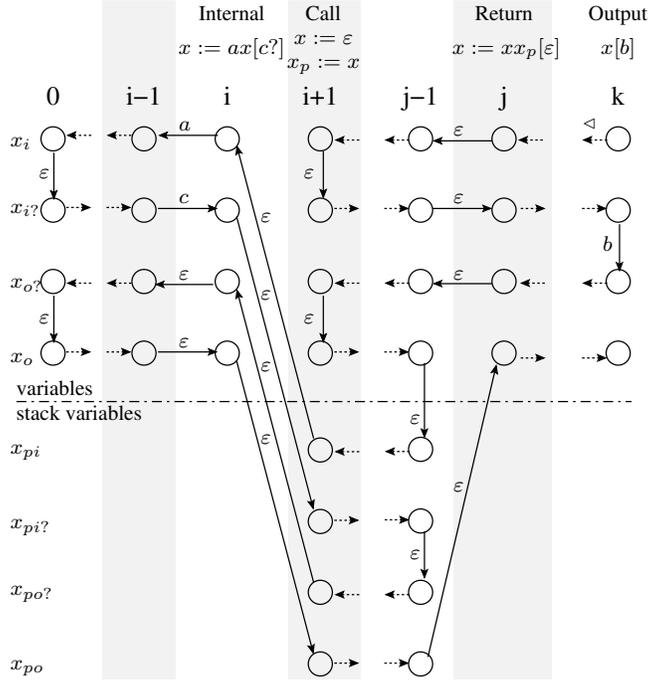}
\caption{Encoding STT computation in MSO}
\label{msoex}
\end{figure}

We explain the encoding using an example shown in Fig. {\ref{msoex}}.
Suppose the STT uses one variable $x$ of type-1.
The corresponding MSO transducer has eight copies in the copy set (four for each variable).
Every variable at every step is represented by 4 nodes in the copy set
called $input$, $output$, $input?$ and $output?$.
At every step $i$
the value of each variable corresponds to the sequence of symbols labeling the unique
path starting at the input copy and ending at the output copy inserting a $?$ labeled link
between the $input?$ and $output?$ nodes.
The value of the variable when stored on the stack ($x_p$ in this example) stored on the top-of-the-stack
at a given step $i$ is similarly captured by the sequence of symbols labeling the unique
path starting at its input copy in column $i$ and ending at the output copy in the same column.

We explain now how variable updates at each step are captured.
Consider an internal position $i$, and consider the variable assignment
$x:=ax[c?]$. This means that the value of $x$ for column $i$
is the value of $x$ in column $i-1$, preceded by the symbol $a$, where we add a $c$
before the parameter position in $i-1$.
To process this
assignment, we insert an $a$-labeled edge from the input node of $x$ in column $i$
to the input node of $x$ in column $i-1$,
we insert an $c$-labeled edge from the $input?$ node of $x$ in column $i-1$
to the $input?$ node of $x$ in column $i$ (this basically shift the parameter position),
we insert an $\varepsilon$-labeled edge from the $output?$ node of $x$ in column $i$
to the $output?$ node of $x$ in column $i-1$ and
we insert an $\varepsilon$-labeled edge from the $output$ node of $x$ in column $i-1$
to the $output$ node of $x$ in column $i$.

To understand how the values of the variables are propagated on the stack, again see Figure~\ref{msoex}.
At the call step $i+1$,
the assignment $x_p:=x$ in which $x$ is stored on the stack is reflected
by the $\varepsilon$-labeled edge from the input node of $x_p$ to input node for $x$. The other edges are similar to before.
Note that the updates until this step do not use $x_p$, and thus, there cannot be any edge
connecting the input/output nodes for $x_p$ in column $i-1$ (where we already have nodes for $x_p$ even though they do not appear in the figure).
In fact, the value of $x_p$ in column $i$ is preserved unchanged until the corresponding matching return
position will be found.
At the return step $j$, the value of $x$ can depend on the values
of $x$ in column $j-1$, and the value of $x_p$ on the top stack, which is captured by
the input/output nodes for $x_p$ in column $j-1$.
Even though it is not shown in figure, at position $j$ we have to add $\varepsilon$ edges from the values of $x_p$
at position $j$ to the values of $x_p$ at position $i$ to represent the value of $x_p$ that now is on the top of the stack.

At step $0$, each variable is instantiated with $\varepsilon$ by adding an
$\varepsilon$-labeled edge from its $input$ node to $input?$ node and from its $output?$ node to $output$ node.


To represent the final output, we need an additional column.
In the example, the output is $x[b]$. So we
mark the first edge of $x$ by a special symbol $\beginstring$ to indicate where the output string
starts and we add
a $b$-labeled edge from the $input?$ node of $x$ to the $output?$ node of $x$. We also have to link it to the previous values with $\varepsilon$ edges.

Note that in this exposition, we have assumed that in the MSO transducer, edges can be labeled
with strings over the output alphabet (including $\varepsilon$) instead of single symbols.
It is easy to show that allowing strings to label the edges of the output graph does not increase the
expressiveness of MSO transducers.
Also note that not every node will appear in the final output string. An MSO transducer able to remove the useless edges and nodes can be defined.
Using closure under composition we can then build the final transducer.

Some extra attention must be paid to add the matching edges in the output. Fortunately the output at every point is always a well-matched word, and so the matching relation is induced by single assignments (the matching edges are always between nodes in the same column).
\qed

\mypar{Nested Words as Binary Trees}
Nested words can be encoded as binary trees.
This encoding is analogous to encoding of {\em unranked\/} trees as binary trees.
Such an encoding increases the depth of the tree by imposing unnecessary hierarchical structure, and
thus, is not suitable for processing of inputs. However, it is useful to simplify proofs of
subsequent results about expressiveness.
The desired transduction $\nwbt$ from $\W_0(\Sigma)$ to $\BT(\Sigma)$ is defined by
\begin{eqnarray*}
\nwbt(\varepsilon)&= &\zero\\
\nwbt(aw)& = & a\lt \,\nwbt(w)\, \zero\,\rt\\
\nwbt(\lt a\,w_1\,b\rt\,w_2)&=&a\lt \,\nwbt(w_1)\, b\lt \, \nwbt(w_2)\,\zero\,\rt\,\rt
\end{eqnarray*}
Note that the tree corresponding to a nested word $w$ has exactly one internal node for
each position in $w$.
Observe that $\nwbt$ is a one-to-one function., and in particular,
the encodings of the two nested words $aa$ and $\lt a\,a\rt$ differ.
We can define the inverse partial function $\btnw$ from binary trees to nested words as follows:
given $t\in\BT(\Sigma)$, if $t$ equals $\nwbt(w)$, for some $w\in\W_0(\Sigma)$
(and if so, the choice of $w$ is unique), then $\btnw(t)=w$, and otherwise $\btnw(t)$ is undefined.
The next proposition shows that both these mappings can be implemented as STTs.

\begin{proposition}[Nested-Words Binary-Trees Correspondence]
The transductions $\nwbt:\W_0(\Sigma)\mapsto \BT(\Sigma)$ and $\btnw:\BT(\Sigma)\mapsto\W_0(\Sigma)$
are STT-definable.
\end{proposition}
\Proof~
We give an idea of how to construct such STTs.
The transition $\nwbt$ can be performed by an STT that basically simulates its inductive definition. We only need one variable $x$.
Every time an $a$ is read we just update $x:=x[\lt a ? \zero a ]\rt$. On input $\lt a$ we store $a$ and $x$ on the stack.
At the corresponding return $b\rt$, $x$ will contain the value of $\nwbt(w_1)$ and so we can update $x:=x_p[\lt a x[\zero] \lt b ? \zero b\rt a\rt]$ and keep reading $w_2$ and its value will be inserted in $x$. The initial value of $x$ is $?$.
The translation $\btnw$ can be implemented as a BRTT in a trivial way.
\qed

For a nested-word transduction $f$ from $\W_0(\Sigma)$ to $\W_0(\Gamma)$,
we can define another transduction $\tilde{f}$ that maps binary trees over $\Sigma$ to
binary trees over $\Gamma$:
given a binary tree $t\in\BT(\Sigma)$,
if  $t$ equals $\nwbt(w)$, then $\tilde{f}(t)=\nwbt(f(w))$, and otherwise $\tilde{f}(t)$ is undefined.
The following proposition can be proved easily from the definitions of the encodings:

\begin{proposition}[Encoding Nested-Word Transductions]
If $f$ is an MSO-definable transduction from $\W_0(\Sigma)$ to $\W_0(\Gamma)$,
then the transduction $\tilde{f}:\BT(\Sigma)\mapsto\BT(\Gamma)$ is an MSO-definable
binary-tree transduction and $f=\btnw\cdot\tilde{f}\cdot\nwbt$.
\end{proposition}

Since STT-definable transductions are closed under composition,
to establish that every MSO-definable transduction is STT-definable,
it suffices to consider MSO-definable transductions from binary trees to binary trees.

\subsection{Macro Tree Transducers}
A Macro Tree Transducer (MTT) \cite{EV85,EM99} is a tree transducer in which
the translation of a tree may not only depend on its subtrees but also on its
context. While the subtrees are represented by input variables, the context information
is handled by parameters.
We refer the reader to \cite{EV85,EM99} for a detailed definition of MTTs, and present here
the essential details.
We only consider deterministic MTTs with regular look ahead
that map binary trees to binary trees.

\def\Trees{{\cal T}}

A {\em (deterministic)
macro-tree transducer with regular look ahead\/} (MTTR) $M$ from $\BT(\Sigma)$
to $\BT(\Gamma)$ consists of
a finite set $Q$ of ranked states,
a list $Y= y_1,\ldots y_n$ of parameter symbols,
variables $X= \{x_l,x_r\}$ used to refer to input subtrees,
an initial state $q_0$,
a finite set $R$ of look-ahead types,
an initial look-ahead type $r_0$,
a look-ahead combinator $\theta : \Sigma\times R \times R \mapsto R$,
and the transduction function $\Delta$.
For every state $q$ and  every look-ahead type  $r$,
$\Delta(q,r)$ is a ranked tree over the alphabet $(Q\times X)\cup \Gamma\cup Y$,
where the rank of a label $(q,x)$ is same as the rank of $q$,
the rank of an output symbol $a\in\Gamma$ is 2,
and the rank of each parameter symbol is 0 (that is, only leaves can be labeled with parameters).

The look-ahead combinator is used to define look-ahead types for
trees: $\theta^*(\zero)=r_0$ and $\theta^*(a\lt\, s_l\, s_r\,\rt)=\theta(a,\theta^*(s_l),\theta^*(s_r))$.
Assume that only the tree $\zero$ has the type $r_0$,
and for every state $q$, $\Delta(q,r_0)$ is a tree over $\Gamma\cup Y$ (the variables $X$ are used to refer
to immediate subtrees of the current input tree being processed, and the type $r_0$ indicates that the input tree has no subtrees).

The MTTR $M$ rewrites the input binary tree $s_0$, and at every step the output tree
is a ranked tree whose nodes are labeled either with an output symbol, or with
a pair consisting of a state of the MTTR along with a subtree of the input tree.
Let $\Trees(s_0)$ denote the set of all subtrees of the input tree $s_0$.
Then, the output $t$ at any step is a ranked tree over $(Q\times \Trees(s_0))\cup \Gamma\cup\{\zero\}$.
The semantics of the MTTR is defined by the derivation relation, denoted by $\Rightarrow$,
over such trees.
Initially, the output tree is a single node labeled with $[ q_0,s_0 ]$.
Consider a subtree of the output of the form $u=[ q, s ](t_1, \ldots t_n)$,
that is, the root is labeled with the state $q$ of rank $n$,
with input subtree $s$, and children of this node
are the output subtrees $t_1,\ldots t_n$. Suppose the look-ahead type of the input subtree $s$ is $r$, and let $s_l$ and $s_r$
be the children of the root.
Let $\chi$ be the tree obtained from the tree $\Delta(q,r)$ by replacing input variables $x_l$ and $x_r$ appearing
in a node label with the input
subtrees $s_l$ and $s_r$ respectively, and replacing each leaf labeled with a parameter $y_l$ by the output subtree $t_l$.
Then, in one step, the MTTR can replace the subtree $u$ with the tree  $\chi$.
The rewriting stops when all the nodes in the output tree are labeled only with output symbols.
That is, for $s\in \BT(\Sigma)$ and $t\in \BT(\Gamma)$,
$\brac{M}(s)=t$ iff $[q_0,s] \Rightarrow^*\, t$.

In general, MTTs are more expressive than MSO.
The restrictions needed to limit the expressiveness rely on the so-called {\em single-use\/} and {\em finite copying}. They enforce an MTT to process every subtree in the input a bounded number of times.
Let $M$ be an MTTR.
\begin{enumerate}
\item
The MTTR $M$ is single use restricted in the parameters (SURP)
if for every state $q$ and every look-ahead type $r$, each parameter $y_j$ occurs
as a node-label at most once in  the tree $\Delta(q,r)$.
\item
The MTTR $M$ is finite-copying in the input (FCI)
if there exists a constant $K$
such that for every tree $s$ over $\Sigma$ and subtree $s'$ of $s$,
if the (intermediate) tree $t$ is derivable from  $[q_0,s]$, then
$t$ contains at most $K$ occurrences of the label $[q,s']$
(and thus, each input subtree is processed at most $K$ times during a derivation).

\end{enumerate}
The following theorem is proved in \cite{EM99}.

\begin{theorem}[Regularity for MTTs]
\label{thm:msomtt}
A ranked-tree transduction $f$
is MSO-definable iff  there exists an MTTR $M$ with SURP/FCI such that $f=\brac{M}$.
\end{theorem}

\subsection{MSO Equivalence}

We first show that bottom-up ranked-tree transducers are as
expressive as MTTs with regular-look-ahead and single-use restriction:

\begin{theorem}[From MTTRs to BRTTs]
If a ranked-tree transduction $f:\BT(\Sigma)\mapsto \BT(\Gamma)$ is definable
by an MTTR with SURP/FCI, then it is BRTT-definable.
\end{theorem}
\Proof~
First of all we notice that in the same way as before we can extend BRTTs to multi-parameter BRTTs. We will consider these ones for sake of clarity.
We are given a MTTR with SURP/FCI  $M=(Q_M,Y_M,q_{0M},R_M,r_{0M},\theta_M,\Delta_M)$.

We divide the proof into several steps and in each of them we use a property of the MTT.
\begin{enumerate}
  \item We compute the transduction
    $f':\BT(\Sigma)\mapsto \BT(R)$ where we replace the input alphabet with its RLA labeling.
This transformation can be expressed as a BRTT.
  \item We compute the function $f'':\BT(R)\mapsto \BT(R')$ where we label each node
of the tree with the set of states in which the MTT processes the corresponding input subtree.
  \item Now that we have the firing sequence we construct a BRTT that computes the function $f''':\BT(R')\mapsto\BT(\Gamma)$. This part relies on the SURP restriction
  \item We then use closure under composition to show that $f=f'\cdot f''\cdot f'''$ is a BRTT definable transformation.
\end{enumerate}

Step 1 is trivial since it just follows the rules of the bottom up automaton. In this step the alphabet $R$ is $R_M\times\Sigma$.
For step 2 we use STTs. STTs can also be viewed as a top down machine and $f''$ is nothing more than a top down relabeling. We now show the construction $S_2$ that implements $f_2$.
We can assume in the following that the MTT is from $\BT(R)$ to $\BT(\Gamma)$. 
We know that
at every point a subtree can be processed by at most $K$ times (the parameter of the FCI) states.
We can label the nodes of the tree with the ordered sequence of states that will process it.
So given a tree over $\BT(R)$ we want to construct a tree
over $\BT(R')$ where $R'=R\times S(Q_M,K)$ and $S(Q_M,K)=\bigcup_{1\leq k\leq K} Q_M^k$ . 

The states of $S_2$ will be over $S(Q_M,K)\cup (S(Q_M,K)\times S(Q_M,K))$. 
The initial state of $S_2$ is $q_{0M}$ (that means the root will be processed only by $q_{0M}$).
The invariant we want to maintain is that whenever we are going to process a left subtree our state will be of the form $(m_1,m_2)$ where $m_1$ is the sequence of states that will process the left subtree and $m_2$ the one that will process the future right subtree. 
When we will start processing the right subtree the state will be $m_2$. So, when processing a left child we store on its stack the state $m_2$ and we will use it at the corresponding return to start processing the right child. At every point the states $m_i$ can be obtained directly from the right hand sides of the rules of the MTT on the sets of states in $m$ (where $m$ is the current state of $S_2$). It's now trivial to do the corresponding labeling using the information stored in the state.

We proceed to step 3. In this step we rely on the SURP property of the MTT.
Notice that processing bottom-up the MTT parameter update, behaves in a different way: if the top down update $y_1:=a(y_2)$ (where $y_1,y_2$  are both representing parameters of some subtree $x$) adds an $a$ on the top of $y_2$. The corresponding bottom up update is $x:=x[y_1\mapsto a(y_2)]$, where the new visible parameter is $y_2$.
We now formalize this idea.

We want to construct a BRTT $S_3=(Q_S,q_{0S},\Pi,X_S,F_S,\delta_S,\rho_S)$ from $\BT(R')\mapsto\BT(\Gamma)$.

The state set $Q_S$ and the transition function $\delta_S$ are defined to capture the same language on which $M$ is defined.
All the control on variables can be inferred from the input alphabet. In the case of total functions one state will be enough.

Thanks to the FCI restriction we know that each subtree will be processed at most in $K$ possible ways (for some $K$).
$X_S$ contains $K$ variables, $\{x_1,\ldots,x_{K}\}$, that after processing a subtree will contain the values of its $K$ possible computations in M. At the beginning all the variable values are  set to $\varepsilon$. 
Our parameter set will be $\Pi=Y_M$. Since the MTT is SURP, at every point, any variable can contain at most one occurrence of each $y_i\in Y_M$.

Now we define the update functions ok $S_3$.
Let's start from the leaf rules.
Let's assume the current leaf is labeled with a sequence $m=q_1\ldots q_j$ and RLA state $r$. For every $q_i\in m$ such that
$\Delta(q_i,r)=t_i(y_1,\ldots,y_k)$ (we can assume without loss of generality that all the states have exactly $k$ parameters) we update
$x_{i}:=t_i(y_1,\ldots,y_k)$ where $y_1,\ldots,y_k$ are parameters. Since the MTT is SURP our variable will have at most one occurrence of each $y_i$.

We now analyze the general rules.
Let's assume the node we are processing is labeled with a sequence $m=q_1\ldots q_j$ and RLA state $r$. For every $q_i\in m$
$\Delta(q_i,r)$ will be of the form
$t_i(Y,(q_{1,1}^i,x_1),\ldots,(q_{1,a_i}^i,x_1),(q_{2,1}^i,x_2),\ldots,(q_{2,b_i}^i,x_2))$, where 
$q_{1,1}^i\ldots q_{1,a_i}^i$ is the sequence of node processing the left subtree
while ,
$q_{2,1}^i\ldots q_{2,b_i}^i$ is the sequence of node processing the right subtree.
By the construction of $S_2$, the left child (and similarly the right) must have been labeled with the sequence
$m_l=q_{1,1}^1\ldots q_{1,a_1}^1\ldots q_{1,1}^j\ldots q_{1,a_j}^j$ such that $|m_l|\leq K$.
Moreover we will have that for all $x_i\in X_l$ (similiarly for $X_r$), $x_i$ will contain the output of $M$ when processing the left child of the 
current node starting in state $q_s$ where $q_s$ is the $s$-th element of the sequence $m_l$ (assuming the parameter are not instantiated yet).
Now we have all the ingredients to complete the rule. The right hand side of a variable $x_i$ will contain
the update corresponding to the rule in $M$ where we replace every state with the corresponding variable in the linearization stated above
and parameters are updated via substitution.
We need to define the conflict relation $\eta$. Not surprisingly the transition relation defined above is copyless, so the reflexive relation
will be enough. The output function $F_S$ will simply output $x_1$, the transformation of the input tree starting in $q_{0M}$.

In step 4 we use closure under composition to create the final STT. This completes the proof.
\qed

Now, we can put together all the results to obtain the main result:

\begin{theorem}[MSO Equivalence]
A nested-word transduction $f:\W_0(\Sigma)\mapsto \W_0(\Gamma)$ is STT-definable
iff it is MSO-definable.
\end{theorem}

\section{Decision Problems}

In this section, we show that a number of analysis problems for our model
are decidable.

\subsection{Output Analysis}

Given an input nested word $w$ over $\Sigma$, and an STT $S$ from $\Sigma$ to $\Gamma$,
consider the problem of computing the output $\brac{S}(w)$.
To implement the operations of the STT efficiently, we can store
the nested words corresponding to variables in linked lists with reference variables pointing to positions
that correspond to holes.
To process each symbol in $w$, the copyless update of variables
can be executed by changing only a constant number of pointers.

\begin{proposition}[Computing Output]
Given an input nested word $w$ and an STT $S$, the output word  $\brac{S}(w)$
can be computed in time $\bigO(|w|)$.
\end{proposition}

The second problem we consider corresponds to {\em type checking}:
given regular languages $L_{pre}$ and $L_{post}$ of nested words over $\Sigma$,
and an STT $S$ from $\Sigma$ to $\Gamma$,
the type checking problem is to determine if $\brac{S}(L_{pre})\subseteq L_{post}$ (that is,
if for every $w\in L_{pre}$, $\brac{S}(w)\in L_{post}$).

\begin{theorem}[Type-Checking]
\label{thm:type-check}
Given an STT $S$ from $\Sigma$ to $\Gamma$, an NWA $A$ accepting nested words over $\Sigma$,
and an NWA $B$ accepting nested words over $\Gamma$,
checking $\brac{S}(L(A))\subseteq L(B)$
is solvable in time $\bigO(|A|^3\cdot |S|^3\cdot n^{kn^2})$ where $n$ is the number of states of $B$, and
$k$ is the number of variables in $S$.
\end{theorem}
\Proof~
The construction is similar to the one of closure under composition.
From $S$, $A$, and $B$, we construct an NWA $P$ that accepts  a nested word $w$ exactly when
$w$ is accepted by $A$ but $\brac{S}(w)$ is not accepted by $B$.
The states of $P$ are triplets $(q_A,q_S,f)$
where $q_A$ keeps track of the state of $A$,
$q_S$ the state of $S$, and $f$ is a function that, for every
variable $x$ of $S$ and states $q_1,q_2$ in $B$, $f(x,q_1,q_2)=(q_1',q_2')$
maps $x,q_1,q_2$ to a pair of states of $B$ $(q_1',q_2')$ such that
there is an execution in $B$ from $q_1$ to $q_1'$ on the word contained in $x$ on the left of $\hole$
and there is an execution on $B$ from $q_2$ to $q_2'$ on the output word contained in $x$ on the right of $\hole$ assuming we use the stack produced from the left part.
The final states of the machine are those where $A$ is final and the summary of the output leads to a non accepting state in $B$.
\qed

As noted in Proposition~\ref{prop:img-reg}, the image of an STT is not necessarily regular.
However, the pre-image of a given regular language is regular, and can be computed.
Given an STT $S$ from input alphabet $\Sigma$ to output alphabet $\Gamma$,
and a language $L\subseteq \W_0(\Gamma)$ of output words,
the set $\preimg(L,S)$ consists of input nested words $w$ such that $\brac{S}(w)\in L$.

\begin{theorem}[Computing Pre-Image]\label{thm:preimage}
Given an STT $S$ from $\Sigma$ to $\Gamma$, and an NWA $B$ over $\Gamma$,
there is an algorithm to compute an NWA $A$ over $\Sigma$ such that $L(A)=\preimg(L(B),S)$.
\end{theorem}
\Proof~
The proof follows from closure under composition. Let's consider $B$ as an STT.
Now we can compute $S'$ as the composition of $S$ and $B$. It doesn't take too long to convince ourselves that
$S'$ considered as an acceptor is exactly $A$.
\qed

It follows that given an STT $S$ and a regular language $L$ of output nested words,
there is an {\sc ExpTime} algorithm to test whether $\img(S)\cap L$ is non-empty.

\subsection{Functional Equivalence}

Finally, we consider the problem of checking {\em functional equivalence\/} of two STTs:
given two streaming tree transducers $S$ and $S'$, we want to check
if they define the same transduction.
Given two {\em streaming string transducers} $S$ and $S'$, \cite{AC10,AC11} shows how to construct
an NFA $A$ over the alphabet $\{0,1\}$ such that the two transducers are {\em inequivalent\/}
exactly when $A$ accepts some word $w$ such that $w$ has equal number of $0$'s and $1$'s.
The idea can be adopted for the case of STTs, but $A$ now will be a nondeterministic pushdown automaton.
The size of $A$ is polynomial in the number of states of the input STTs, but exponential in
the number of variables of the STTs.
Results in \cite{Esp97,SSMH04} can be adopted to check whether this pushdown automaton
accepts a word with the same number of
$0$'s and $1$'s.

\begin{theorem}[Checking Equivalence]\label{thm:equivalence}
Given two STTs $S$ and $S'$, the problem of checking whether $\brac{S}\not=\brac{S'}$ is
solvable in {\sc NExpTime}.
\end{theorem}
\Proof~
Two streaming tree transducers $S$ and $S'$ are inequivalent if either:
\begin{enumerate}
  \item for some input $u$ only one of $\brac{S}(u)$ and $\brac{S'}(u)$ is defined or
  \item for some input $u$ the lengths of $\brac{S}(u)$ and $\brac{S'}(u)$ differ or
  \item for some input $u$ there exist two symbols $a,b$ such that $a\not =b$ and $\brac{S}(u)=u_1 au_2$ and $\brac{S'}(u)=v_1 bv_2$ such that
    $u_1$ and $v_1$ have the same length.
\end{enumerate}

The first two cases can be checked with lower complexity. The first one as shown in \cite{AM09} is in {\sc PTime}, and the second one can be reduced to checking an affine relation over PDA that in \cite{Muller-Olm:2004:PIA:982962.964029} is proven to be {\sc PTime}.


Let us focus on (the more interesting) case 3) in which the outputs differ in some position.
Given $S$ and a symbol $a$ we construct a nondeterministic visibly pushdown transducer (a visibly pushdown automata with output) $V_1$ from $\Sigma$ to $\{0\}$
such that $0^n$ is produced by $V_1$ if for some $u$, $\brac{S}(u)=u_1 au_2$ and $|u_1|=n$.

The states of $V_1$ are pairs $(q,f)$ where $q$ is a state of $S$ and $f$ is a
partition of the variables $X$ of $S$ into 6 categories:
\emph{l)} the variable contributes to the final output occurring on the left of a symbol $a$ where $a$ is the symbol we have guessed the two transducers differ in the final output,
\emph{m1)} the variable contributes to the final output and the symbol $a$ appears in this variable on the left of the $\hole$,
\emph{m?)} the variable contributes to the final output and the symbol $a$ will appear in this variable in the $\hole$ (a future substitution will add $a$ to the $\hole$,
\emph{m2)} the variable contributes to the final output and the symbol $a$ appears in this variable on the right of the $\hole$,
\emph{r)} the variable contributes to the final output occurring on the right of a symbol $a$,
\emph{n)} the variable does not contribute to the final output.

At every step, $V_1$ nondeterministically chooses which of the previous categories
each of the variables of $S_1$
belongs to.
In the following we denote as $f_i$ the partitions defined before (i.e. given $f$, $f_{m1}$ is the set of variables mapped to $mq$).
A state $(q,f)$ is initial in $V_1$ if $q$ is an initial state in $S$, and $f_{m1}\cup f_{m2}=\emptyset$.
A careful reader will notice that an STT doesn't have final states but we can get rid of this problem creating a final state $q_f$ and adding a *-transition from all the states in which the output function is defined using a symbol $*$ to label the transition where $*\not\in\Sigma$.
We still have the problem of which
variable will contain the output and we can solve it updating $x$ (the first variable) to the value of the output function.
Let's refine the definition then: a state $(q,f)$ is final in $V_1$ if $q=q_f$, $f_{m1}=\{x\}$ (notice that at this point the variable can't contain parameters and it has to be of type-0, so we do not need to consider $f_{m2}$) and $f_n=X\setminus \{x\}$
(the only variable contributing to the output is $x$). Clearly $f_l\cup f_r \cup f_{m?}\cup f_{m2}=\emptyset$.

Transitions of $V_1$ ensure that these attributes are consistently updated. We now explain how they work formally.
Given $(q,f)$ on input $s$ we have the following possibilities (we denote by $f_j$ with $j\in\{l,m1,m?,m2,r,n\}$ the corresponding partition,
and given a string $\alpha$ we say that a variable $x\in \alpha$ if it occurs in it):

\mypar{$s$ is internal} $(q,f)$ steps to $(q',f')$ where $\delta_i(q,s)=q'$.
    To update $f$ we have 3 possible cases:\\
    i) we guess that in this transition, some variable $x$ is going to contain the guessed position containing the symbol on which the output differ,\\
    ii) the transition is just maintaining the consistency of the partition and the position on which
    the output differs hasn't been guessed yet,\\
    iii) the transition is just maintaining the consistency of the partition and the position on
    which the output differs has already been guessed.

    \emph{Case i):} let's assume the guess is that $\rho_i(q,s,x)=\alpha_1 a \alpha_2 ? \alpha_3$ and $a$
    is the position on which we guess the output differs.
    To perform a consistent update we need the transition to satisfy the following properties:
    $\forall y\in \alpha_1. y\in f_l$, $\forall y\in \alpha_2\alpha_3. y\in f_r$, $f'_{m1}=\{x\}$ and $f_m=\emptyset$
    (the only variable that contributes in the middle now is $x$),
    given a variable $y\not =x$ all the variables in $\rho_i(q,s,y)$ belong to the
    same partition $f_j$ and $y\in f'_j$. If a variable is assigned a constant we nondeterministically choose
    which category it will belong to in $f'$ (we omit this detail in next points).
    In this case the output is $0^k$ where $k$ is the sum of the number of input
    symbols in $\alpha$ and in $\{\rho_i(q,s,y)| y\in f'_l\}$.
    Some extra hack is needed for assignments where we do parameter substitution: $\rho_i(q,s,x)=x[a]$. In this
    we better have guessed that $f_{m?}=\{x\}$ and $f'_{m1}=\{x\}$.

    \emph{Case ii and iii):} Similar to before.


\mypar{$s$ is a call} in this case the updates are similar with the difference that we have to store on
      the stack a state that records the partition of the variables at the call.
      Reading $s$, $(q,f)$ steps to $(q',f')$, pushes $(p,f'')$ where $\delta_c(q,s)=q',p$. $f''$ will be the updated partition talking about the variables in $X_p$. $f'$ is a new partition for the reset variables.

\mypar{$s$ is a return} returns ar a bit more interesting than calls since we have to deal with the previous value of the variables stored on the stack, but still the definition is the same of that for internal action.

      Reading $s$, $(q,f)$ with $(p,f')$ on top of the stack, steps to $(q'',f'')$ where $\delta_r(q,p,s)=q'$.
      We show how the first case differs:
      let's assume the guess is that for a variable $x\in X$, $\rho_r(q,s,x)=\alpha a \alpha' ? \alpha''$ and $a$ is the
      position on which the output differs.
      To perform a consistent update we need the transition to satisfy the following properties:
      $\forall y\in \alpha. y\in f_l\cup f'_l$, $\forall y\in \alpha'\alpha''. y\in f_r\cup f'_r$,
      $f''_{m1}=\{x\}$, $f_{m1}\cup f_{m?}\cup f_{m2}=\emptyset$ and $f'_{m1}\cup f'_{m?}\cup f'_{m2}=\emptyset$,
      given a variable $y\not =x$ all the variables in $\rho_r(q,p,s,y)$ are in $f_j\cup f'_j$ and $y\in f''_j$ (for some $j\in \{l,r,n\}$).
      In this case the output is $0^k$ where $k$ is the sum of the number of
      input symbols in $\alpha$ and
      in $\{\rho_r(q,p,s,y)| y\in f''_l\}$.

We actually impose the extra condition on transitions that the cardinality of $f_{m1}\cup f_{m2}$ is always less or equal than $1$ since at most one variable
can contain the symbol on which the output differs. Moreover another condition is that if a variable doesn't appear in the right hand side of any
assignment it should be in $f_n$.

Then given $S'$ and a symbol $b\not =a$ we construct a nondeterministic VPT $V_2$ from $\Sigma$ to $\{1\}$
such that $1^n$ is produced by $V_2$ if for some $u$, $S(u)=u_1 bu_2$ and $|u_1|=n$.

Now we take the product $V=V_1\times V_2$.
Once we take the product, input labels are no longer relevant,
and we can view it as a pushdown automaton that generates/accepts strings over $\{0,1\}$.

We want to check if $V$ accepts some string that contains the same number of $0$'s and $1$'s
(which would ensure that the number of symbols contributed by $S_1$ to the left of $a$
equals the corresponding number for $S_2$ to the left of $b$).
This can be solved by constructing the semi-linear set that characterizes the Parikh image
of the context-free language of $V$~\cite{Esp97,SSMH04}, and can be solved
in {\sc NP} (in the number of states of $V$).

The number of states of $V$ is polynomial in the number of states of the transducers
$S$ and $S'$, but exponential in the number of variables of the transducers
(due to the classification of each variable into 6 different categories).
This gives the bound {\sc NExpTime} for the inequivalence check.
\qed

If the number of variables is bounded, then the size of $V$ is polynomial, and
this gives an upper bound of {\sc NP}.
For the transducers that map strings to nested words,
that is, for streaming string-to-tree transducers (SSTT),
the above construction yields a {\sc Pspace} bound:

\begin{theorem}[Equivalence of String-to-tree Transducers]
Given two SSTTs $S$ and $S'$ that map strings to nested words,
the problem of checking whether $\brac{S}=\brac{S'}$ is
solvable in {\sc Pspace}.
\end{theorem}

\section{Discussion}
We have proposed the model of streaming tree transducers to implement
MSO-definable tree transformations by processing the linear encoding of the input tree
in a single left-to-right pass in linear time.
Below we discuss the relationship of our model to the rich variety of existing transducer models,
and directions for future work.

\mypar{Executable models}
A streaming tree transducer is an executable model, just like a deterministic automaton
or a sequential transducer, meaning that the operational semantics of the machine
processing the input coincides with the algorithm to compute the output from the input
and the machine description.
Earlier executable models for tree transducers include
bottom-up tree transducers, visibly pushdown transducers
(a VPT is a sequential transducer with a visibly pushdown store: it reads the input nested word
left to right producing output symbols at each step)~\cite{RS08},
and multi bottom-up tree transducers (such a transducer computes a bounded number of transformations
at each node by combining the transformations of subtrees)~\cite{ELM08}.
Each of these models computes the output in a single left-to-right pass in linear time.
However, none of these models  can compute all MSO-definable transductions,
and in particular, can compute the transformations such as swap and tag-based sorting.

\mypar{Regular look ahead}
Finite copying Macro Tree Transducers (MTTs) with regular look ahead can compute
all MSO-definable ranked-tree-to-ranked-tree transductions.
The ``finite copying'' restriction, namely, each input node is processed
only a bounded number of times, can be equivalently replaced by
the syntactic ``single use restriction'' which restricts how the
variables and parameters are used in the right-hand sides of rewriting rules
in MTTs.
In all these models, regular look ahead cannot be eliminated without sacrificing
expressiveness: all of these process the input tree in a top-down manner, and it is well-known
that deterministic top-down tree automata cannot specify all tree regular languages.
A more liberal model with ``weak finite copying'' restriction
achieves closure under regular look ahead, and MSO-equivalence, by allowing
each input node to be processed an unbounded number of times, provided only a bounded
subset of these contribute to the final output.
It should be noted, however, that a linear time algorithm exists to compute the output~\cite{Man02}.
This algorithm essentially uses
additional look ahead passes to label the input with the information needed to restrict attention to only
those copies that will contribute to the final output (in fact,
\cite{Man02} shows how relabeling of the input can be effectively used to compute the output
of every MTT in time linear in the size of the input and the output).
Finally, to compute tree-to-string transductions, in presence of regular look ahead,
MTTs need just one parameter (alternatively, top-down tree transducers suffice).
In absence of regular look ahead, even if the final output is a string, the MTT needs multiple
parameters, and thus, intermediate results must be trees
(that is, one parameter MTTs are not closed under regular look ahead).
Thus, closure under regular look ahead is a key distinguishing feature of STTs.

\mypar{From SSTs to STTs}
The STT model generalizes our earlier work on streaming string transducers (SST):
SST is a copyless STT without a stack~\cite{AC10,AC11}.
 While results in Section~5 follow by a natural
generalization of the corresponding results for SSTs, the results in Section~3 and 4
require new approach. In particular, equivalence of SSTs with MSO-definable string-to-string
transductions is proved by simulating a two-way deterministic sequential transducer,
a well-studied model known to be MSO-equivalent~\cite{EH01}, by an SST.
The MSO-equivalence proof in this paper first establishes closure under regular look
ahead, and then simulates finite copying MTTs with regular look ahead.
The natural analog of two-way deterministic string transducers would be the two-way
version of visibly pushdown transducers~\cite{RS08}: while such a model has not been studied,
it is easy to show that it would violate the ``linear-bounded output'' property of Proposition~1,
and thus, won't be MSO-equivalent.

\mypar{Succinctness}
To highlight the differences in how MTTs and STTs compute, we consider two
``informal'' examples.
Let $f_1$ and $f_2$ be two MSO-definable transductions, and
consider the transformation $f(w)=f_1(w)f_2(w)$.
An MTT at every node can send multiple copies to children, and thus, has
inherent parallelism. Thus, it can compute $f$ by having one copy compute $f_1$,
and one copy compute $f_2$, and the size of the resulting MTT will be the sum of the sizes
of MTTs computing $f_1$ and $f_2$.
STTs are sequential, and thus, to compute $f$, one needs the product of the
STTs computing  $f_1$ and $f_2$.
This can be generalized to show that MTTs (or top-down tree transducers)
can be exponentially more succinct than STTs.
If we were to restrict MTT rules so that multiple states processing the same subtree must
coincide, then this gap disappears.
In the other direction, consider the transformation $f'$ that maps
input $u\#v\#a$ to $uv$ if $a=0$ and $vu$ otherwise.
The transduction $f'$ can be easily implemented by an STT using two variables,
one of which stores $u$ and one which stores $v$.
The ability of an STT to concatenate variables in any order allows it to output
either $uv$ or $vu$ depending on the last symbol.
In absence of look ahead, an MTT for $f'$ must use two parameters,
and compute (the tree encodings of) $uv$ and $vu$ separately in parallel,
and make a choice at the end.
This is because, while an MTT rule can swap or discard output subtrees
corresponding to parameters, it cannot combine subtrees corresponding to
parameters.
This example can be generalized to show that an MTT must use exponentially many
parameters as well as states compared to an STT.


\mypar{Input/output encoding}
Most models of tree transducers process ranked trees
(exceptions include visibly pushdown transducers~\cite{RS08} and Macro forest transducers~\cite{PS04}).
While an unranked tree can be encoded as a ranked tree (for example, a word
of length $n$ can be
viewed as a unary tree of depth $n$), this is not a good encoding choice for processing the input,
since the stack height is related to depth (in particular, processing
a word does not need a stack at all).
We have chosen to encode unranked trees by nested words;
formalization restricted to tree words (that are isomorphic to unranked trees)
would lead to a slight simplification of the STT model and the proofs.

\mypar{Streaming algorithms}
Consistent with the notion of a streaming algorithm, an STT processes
each input symbol in constant time.
However, it stores the output in multiple chunks in different variables,
rearranging them without examining them, making decisions based on finite-state control.
Unlike a typical streaming algorithm, or a sequential transducer,
the output of an STT is available only after reading the entire input.
This is unavoidable if we want compute a function that maps an input to its reverse.
We would like to explore if the STT model can be modified so that it commits
to output symbols as early as possible. A related direction of future work
concerns minimization of resources (states and variables).

\mypar{Complexity of checking equivalence}
The problem of checking functional equivalence of MSO tree transducers is decidable with
nonelementary complexity~\cite{EM06}.
Decidability follows for MSO-equivalent models such as MTTs with finite copying,
but no complexity bounds have been established.
Polynomial-time algorithms
for equivalence checking exist for top-down tree transducers
(without regular look ahead) and visibly pushdown transducers~\cite{TATA,RS08,EMS09}.
For STTs, we have established an upper bound of {\sc NExpTime}, while the upper bound
for SSTs is {\sc Pspace}~\cite{AC11}.
Improving these bounds, or establishing lower bounds, remains a challenging open problem.
If we extend the SST/STT model by removing the single-use-restriction on variable updates,
we get a model more expressive than MSO-definable transductions; it remains
open whether the equivalence problem for such a model is decidable.


\mypar{Application to XML processing}
We have argued that SSTs correspond to a natural model with executable interpretation,
adequate expressiveness, and decidable analysis problems, and in future work,
we plan to explore its application to
querying and transforming XML documents~\cite{Ho11} (see also \url{http://www.w3.org/TR/xslt20/}).
Our analysis techniques typically have complexity that is exponential in the number of
variables, but we do not expect the number of variables to be the bottleneck.
Before we start implementing a tool for XML processing, we want to understand how to integrate
data values (that is, tags ranging over a potentially unbounded domain) in our model.
A particularly suitable implementation platform for this purpose seems to be
the framework of {\em symbolic automata\/} and {\em symbolic transducers\/}
that allows integration of automata-theoretic decision procedures on top
of the SMT solver Z3 that allows manipulation of formulas specifying input/output values
from a large or unbounded alphabet in a symbolic and succinct manner~\cite{BV12}.

\mypar{Acknowledgments}
We thank Joost Engelfriet for his valuable feedback:
not only he helped us navigate the extensive literature on tree transducers,
but also provided detailed comments, including spotting bugs in proofs,
on an earlier draft of this paper.

\bibliographystyle{abbrv}
\bibliography{stt12}

\begin{thebibliography}{10}

\bibitem{TATA}
Comon, H., Dauchet, M., Gilleron, R., Lugiez, D., Tison, S., Tommasi, M.:
\newblock Tree automata techniques and applications.
\newblock Draft, Available at \verb+http://www.grappa.univ-lille3.fr/tata/+
  (2002)

\bibitem{Cou94}
Courcelle, B.:
\newblock Monadic second-order definable graph transductions: A survey.
\newblock Theor. Comput. Sci. \textbf{126}(1) (1994)  53--75

\bibitem{EM99}
Engelfriet, J., Maneth, S.:
\newblock Macro tree transducers, attribute grammars, and {MSO} definable tree
  translations.
\newblock Information and Computation \textbf{154} (1999)  34--91

\bibitem{EV85}
Engelfriet, J., Vogler, H.:
\newblock Macro tree transducers.
\newblock J. Comput. System Sci. \textbf{31} (1985)  71--146

\bibitem{MSV02}
Milo, T., Suciu, D., Vianu, V.:
\newblock Typechecking for xml transformers.
\newblock In: Proceedings of the 19th ACM Symposium on PODS. (2000)  11--22

\bibitem{HP03}
Hosoya, H., Pierce, B.C.:
\newblock X{D}uce: A statically typed {XML} processing language.
\newblock ACM Trans. Internet Techn. \textbf{3}(2) (2003)  117--148

\bibitem{MN05b}
Martens, W., Neven, F.:
\newblock On the complexity of typechecking top-down {XML} transformations.
\newblock Theor. Comput. Sci. \textbf{336}(1) (2005)  153--180

\bibitem{Ho11}
Hosoya, H.:
\newblock Foundations of {XML} Processing: The Tree-Automata Approach.
\newblock Cambridge University Press (2011)

\bibitem{SV02}
Segoufin, L., Vianu, V.:
\newblock Validating streaming {XML} documents.
\newblock In: Proceedings of the 21st ACM Symposium on PODS. (2002)  53--64

\bibitem{NS02}
Neven, F., Schwentick, T.:
\newblock Query automata over finite trees.
\newblock Theor. Comput. Sci. \textbf{275}(1-2) (2002)  633--674

\bibitem{MV09}
Madhusudan, P., Viswanathan, M.:
\newblock Query automata for nested words.
\newblock In: Mathematical Foundations of Computer Science 2009, 34th
  International Symposium. LNCS 5734 (2009)  561--573

\bibitem{AM09}
Alur, R., Madhusudan, P.:
\newblock Adding nesting structure to words.
\newblock Journal of the {ACM} \textbf{56}(3) (2009)

\bibitem{AC10}
Alur, R., Cern{\'y}, P.:
\newblock Expressiveness of streaming string transducers.
\newblock In: IARCS Annual Conference on Foundations of Software Technology and
  Theoretical Computer Science. LIPIcs 8 (2010)  1--12

\bibitem{AC11}
Alur, R., Cern{\'y}, P.:
\newblock Streaming transducers for algorithmic verification of single-pass
  list-processing programs.
\newblock In: Proceedings of 38th ACM Symposium on POPL. (2011)  599--610

\bibitem{EM06}
Engelfriet, J., Maneth, S.:
\newblock The equivalence problem for deterministic {MSO} tree transducers is
  decidable.
\newblock Inf. Process. Lett. \textbf{100}(5) (2006)  206--212

\bibitem{SSMH04}
Seidl, H., Schwentick, T., Muscholl, A., Habermehl, P.:
\newblock Counting in trees for free.
\newblock In: Automata, Languages and Programming: 31st International
  Colloquium. LNCS 3142 (2004)  1136--1149

\bibitem{Esp97}
Esparza, J.:
\newblock Petri nets, commutative context-free grammars, and basic parallel
  processes.
\newblock Fundam. Inform. \textbf{31}(1) (1997)  13--25

\bibitem{EH01}
Engelfriet, J., Hoogeboom, H.:
\newblock {MSO} definable string transductions and two-way finite-state
  transducers.
\newblock ACM Trans. Comput. Log. \textbf{2}(2) (2001)  216--254

\bibitem{Muller-Olm:2004:PIA:982962.964029}
M\"{u}ller-Olm, M., Seidl, H.:
\newblock Precise interprocedural analysis through linear algebra.
\newblock SIGPLAN Not. \textbf{39} (2004)  330--341

\bibitem{RS08}
Raskin, J., Servais, F.:
\newblock Visibly pushdown transducers.
\newblock In: Automata, Languages and Programming: Proceedings of the 35th
  ICALP. LNCS 5126 (2009)  386--397

\bibitem{ELM08}
Engelfriet, J., Lilin, E., Maletti, A.:
\newblock Extended multi bottom-up tree transducers.
\newblock In: Developments in Language Theory. LNCS 5257 (2008)  289--300

\bibitem{Man02}
Maneth, S.:
\newblock The complexity of compositions of deterministic tree transducers.
\newblock In: FST TCS 2002: Foundations of Software Technology and Theoretical
  Computer Science, 22nd Conference. LNCS 2556 (2002)  265--276

\bibitem{PS04}
Perst, T., Seidl, H.:
\newblock Macro forest transducers.
\newblock Inf. Process. Lett. \textbf{89}(3) (2004)  141--149

\bibitem{BV12}
Bjorner, N., Hooimeijer, P., Livshits, B., Molner, P., Veanes, M.:
\newblock Symbolic finite state transducers, algorithms, and applications.
\newblock In: Proc. 39th ACM Symposium on POPL. (2012)

\end{thebibliography}
\end{document}